\documentclass[9pt,twocolumn,twoside]{pnas-new}

\templatetype{pnasresearcharticle} 

\usepackage{graphicx}
\usepackage{amsmath}
\usepackage{color}
\usepackage{verbatim}
\usepackage{bm}
\usepackage{cancel}
\usepackage{ulem}
\usepackage{subfigure}

\newcommand{\be}{\begin{equation}}
\newcommand{\ee}{\end{equation}}
\newcommand{\ba}{\begin{eqnarray}}
\newcommand{\ea}{\end{eqnarray}}

\begin{document}

\title{Fluctuation theorems for autonomous work}

\author[a,b,c,1]{Christopher Jarzynski}
\author[d,e,f]{Sebastian Deffner}
\author[g]{Saar Rahav}

\affil[a]{Department of Chemistry and Biochemistry, University of Maryland, College Park, MD 20742 USA}
\affil[b]{Institute for Physical Science and Technology, University of Maryland, College Park, MD 20742 USA}
\affil[c]{Department of Physics, University of Maryland, College Park, MD 20742 USA}
\affil[d]{Department of Physics, University of Maryland, Baltimore County, Baltimore, MD 21250, USA}
\affil[e]{Quantum Science Institute, University of Maryland, Baltimore County, Baltimore, MD 21250, USA}
\affil[f]{National Quantum Laboratory, College Park, MD 20740, USA}
\affil[g]{Schulich Faculty of Chemistry, Technion - Israel Institute of Technology, Haifa 3200003, Israel}

\leadauthor{Jarzynski}

\significancestatement{Physical systems exchange energy with one another. The second law of thermodynamics constrains these exchanges. In stochastic thermodynamics, fluctuation theorems extend the second law to the regime of small systems. To date, classical fluctuation theorems have been derived within an idealized framework in which systems are manipulated by external agents. We obtain key fluctuation theorems within a different framework, involving systems exchanging energy autonomously, without external interference. Our fluctuation theorems account for the back-action experienced by a thermodynamic work source, as it swaps energy with a system of interest. If the work source has infinite inertia, then the fluctuation theorems we obtain become equivalent to their agent-driven counterparts. Our results reveal that fluctuation theorems apply more broadly than previously appreciated.}

\authorcontributions{Author contributions: C.J, S.D. and S.R. performed research and wrote the paper.}
\authordeclaration{The authors declare no conflict of interest.}
\correspondingauthor{\textsuperscript{1}To whom correspondence should be addressed. E-mail: cjarzyns@umd.edu}

\keywords{fluctuation theorems $|$ second law of thermodynamics $|$ nonequilibrium processes }

\begin{abstract}
Classical fluctuation theorems for work have been obtained theoretically, and verified experimentally, within a non-autonomous framework in which work is performed on a system of interest, ${\cal S}$, by the external manipulation of a \textit{work parameter}, such as a piston's position.
Here we obtain fluctuation theorems within an autonomous framework in which ${\cal S}$ exchanges energy with a \textit{reversible work source}, ${\cal R}$.
The two subsystems, ${\cal R}$ and ${\cal S}$, interact with one another as they evolve under Hamiltonian or stochastic dynamics, without external intervention.
In this setting, we must account for the back-action of ${\cal S}$ on ${\cal R}$, which is absent in the non-autonomous setting.
We obtain autonomous versions of standard fluctuation theorems for work and entropy production.
In each case, we argue, the autonomous fluctuation theorem reduces to its non-autonomous counterpart when ${\cal R}$'s inertia becomes infinitely large.
\end{abstract}

\dates{This manuscript was compiled on \today}
\doi{\url{www.pnas.org/cgi/doi/10.1073/pnas.2524775122}}

\maketitle
\thispagestyle{firststyle}
\ifthenelse{\boolean{shortarticle}}{\ifthenelse{\boolean{singlecolumn}}{\abscontentformatted}{\abscontent}}{}

\begin{figure}[h]
   \subfigure[]{
   \label{fig:aut}
   \includegraphics[trim = 1.5in 2in .9in .9in , scale=0.20,angle=0]{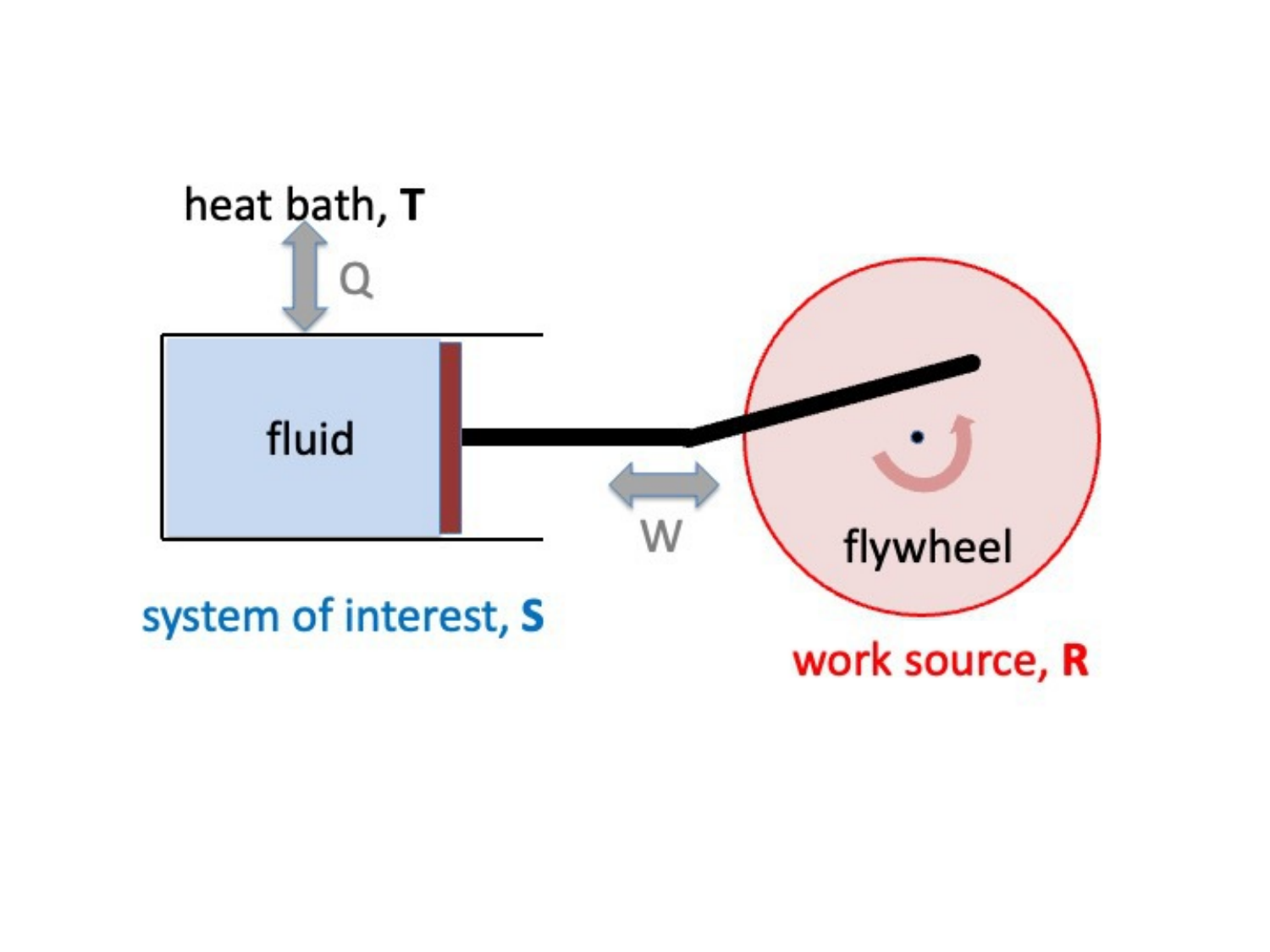}
   }
   \subfigure[]{
   \label{fig:non}
   \includegraphics[trim = 1in 2in .5in .7in , scale=0.20,angle=0]{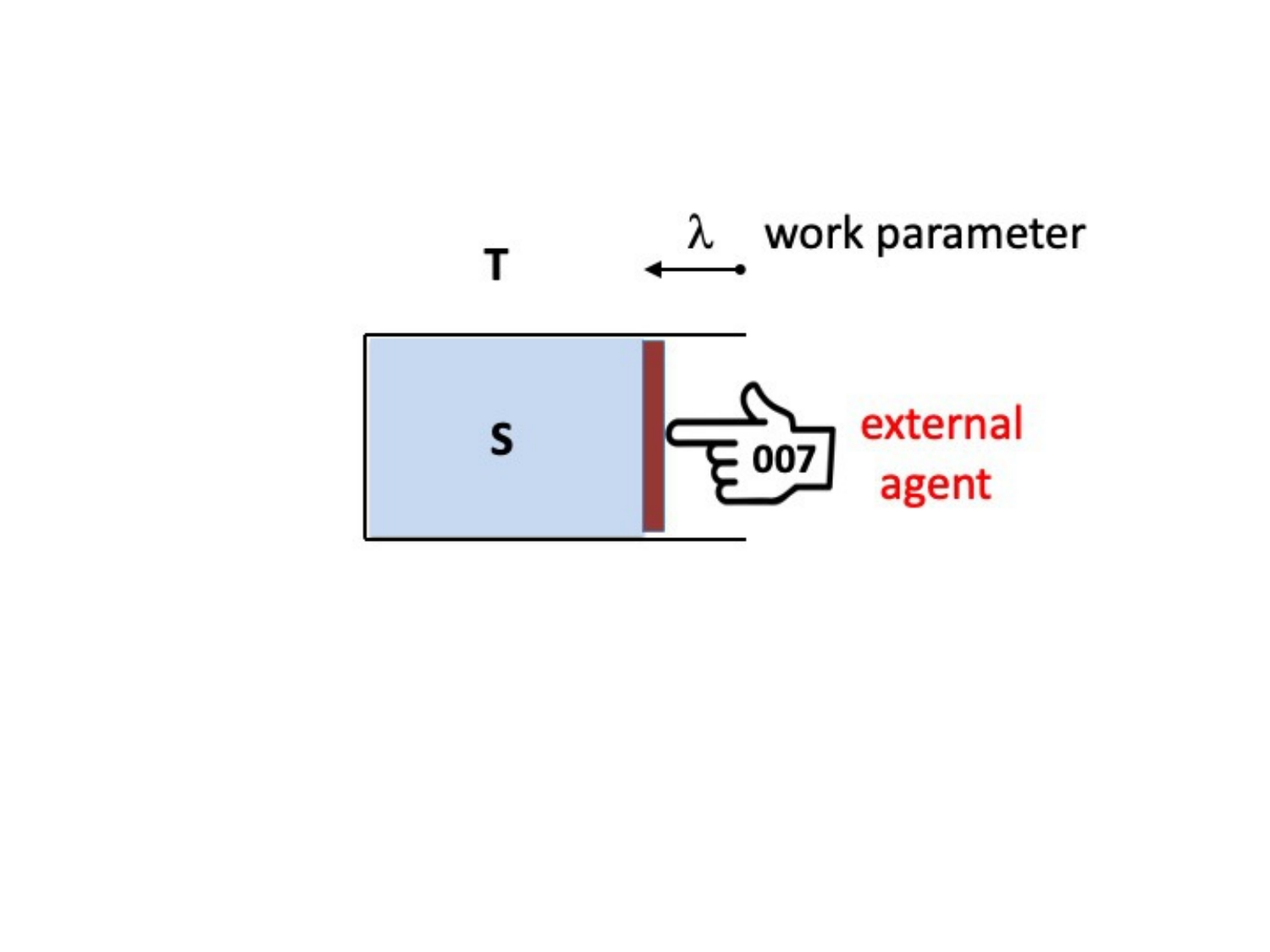}
   }
\caption{(a) {\it Autonomous work}.
A fluid, ${\cal S}$, exchanges energy with a frictionless flywheel, ${\cal R}$.
The flywheel's rotation couples to a piston that acts on the fluid.
As ${\cal R}$ rotates, it performs positive or negative work on ${\cal S}$, losing or gaining rotational kinetic energy in the process.
(b) {\it Non-autonomous work}.
An external agent performs work on the fluid by varying the piston's position, $\lambda$.
The work performed on ${\cal S}$ is $\int F\, d\lambda$, where $F$ is the force that the fluid exerts against the piston.
}
\label{fig:global-1d}
\end{figure}

\dropcap{I}n classical thermodynamics we encounter two distinct notions of work.
(1) Fundamentally, work is the transfer of energy between a system of interest, ${\cal S}$, and a body, or {\it reversible work source}, ${\cal R}$~\cite{Gibbs1902,Callen1985,Anacleto2010}.
This notion of work is illustrated by the fluid ($\cal S$) and flywheel ($\cal R$) in Fig.~\ref{fig:aut}.
We call such work {\it autonomous}, since the energy transfer arises from the dynamics of physical systems evolving without external intervention.
(2) Alternatively, we often see work written as the integral, $\int F\,d\lambda$, of a force $F$ along displacements of a parameter, $\lambda$~\cite{Landau-Lifshitz-StatPhys,Finn1993}, such as the piston's position in Fig.~\ref{fig:non}.
In this notion of work, which we call {\it non-autonomous}, there is no explicit reference to the physical body ${\cal R}$.
In its place, it is convenient to imagine an external agent who controls how the {\it work parameter}, $\lambda$, varies with time.
For macroscopic systems the laws of thermodynamics apply to both autonomous and non-autonomous work, and there is little need to distinguish between the two notions.

{\it Stochastic thermodynamics}~\cite{Seifert2012,Peliti2021,Limmer2024,Seifert2025} examines how thermodynamic laws scale down to small systems.
In this setting, fluctuations in thermodynamic quantities become salient, and the inequalities that express the macroscopic second law are replaced by stronger equalities, or {\it fluctuation theorems}.
Among these results, the nonequilibrium work relation,
\be
\label{eq:ft-na}
\left\langle e^{-\beta W} \right\rangle = e^{-\beta \Delta F } \,,
\ee
relates fluctuations in the work, $W$, performed on a system, to an equilibrium free energy difference, $\Delta F$, where $\beta=T^{-1}$ is an inverse temperature (setting Boltzmann's constant $k_B=1$) and $\langle\cdot\rangle$ denotes an ensemble average~\cite{Jarzynski1997a,Jarzynski1997b}.
Specifically, $\Delta F$ is the free energy difference between the two equilibrium states that correspond to initial and final values of a work parameter, $\lambda$.
By Jensen's inequality~\cite{Chandler1987-Sec5.5}, Eq.~\ref{eq:ft-na} implies
\be
\label{eq:2lt-na}
\langle W \rangle \ge  \Delta F \, ,
\ee
which expresses the second law of thermodynamics in terms of the mean work performed on a system.

Equation~\ref{eq:ft-na} and related results~\cite{Bochkov1977,Bochkov1981,Crooks1998,Crooks1999,Hummer2001,Seifert2005} have been derived for non-autonomous work.
In this paper we obtain corresponding results for autonomous work --  the transfer of energy between a body, ${\cal R}$, and a system, ${\cal S}$.
Our first main result is an integral fluctuation theorem\footnote{
Following the literature, we distinguish between integral fluctuation theorems, expressed as averaged exponentials (Eq.~\ref{eq:ft-na}), and detailed fluctuation theorems, expressed as ratios of distributions (Eq.~\ref{eq:cft}).},
\begin{equation}
\label{eq:ft-a-intro}
\left\langle e^{-\beta ( W - \Delta F ) - \Delta\phi} \right\rangle = 1 \, ,
\end{equation}
whose immediate consequence is the inequality
\begin{equation}
\label{eq:2lt-a-intro}
\beta \langle W - \Delta F \rangle + \Delta S_{\cal R} \ge 0 \, .
\end{equation}
Here, $\Delta\phi$ and $\Delta S_{\cal R} \equiv \langle \Delta \phi \rangle$ reflect the entropy change of the work source, ${\cal R}$, arising from the fluctuating back-action on ${\cal R}$ due to its interactions with ${\cal S}$.
Equations~\ref{eq:ft-a-intro} and \ref{eq:2lt-a-intro} are autonomous counterparts of Eqs.~\ref{eq:ft-na} and \ref{eq:2lt-na}.

In what follows, we first derive Eq.~\ref{eq:ft-a-intro} for a system and work source that are thermally isolated.
We then extend our derivation to include a thermal reservoir, ${\cal T}$, that exchanges heat with ${\cal S}$.
Next we argue that Eq.~\ref{eq:ft-a-intro} reduces to Eq.~\ref{eq:ft-na} in the limit of a work source with infinite inertia -- that is, when the back-action on ${\cal R}$ becomes negligible.
An analogous argument for generalized second-law inequalities was put forward by two of us in Ref.~\cite{Deffner2013_prx}, within a Hamiltonian approach in which autonomous and non-autonomous work become identical in the limit of infinite inertia.
Finally, we show that three other non-autonomous fluctuation theorems found in the literature -- an early integral fluctuation theorem for {\it exclusive work}~\cite{Bochkov1977,Bochkov1981}, an integral fluctuation theorem for entropy production~\cite{Jarzynski1999a,Seifert2005}, and a detailed fluctuation theorem~\cite{Crooks1999} -- have autonomous counterparts.
Our central results are Eqs.~\ref{eq:ft-a-intro} (derived variously in Eqs.~\ref{eq:ft} and \ref{eq:ft-a-rst}), \ref{eq:ft-exc}, \ref{eq:sft-a} and \ref{eq:cft-a}.


\begin{figure}[h]
\includegraphics[trim = 0in 0.5in 5in 0.5in , scale=0.3,angle=0]{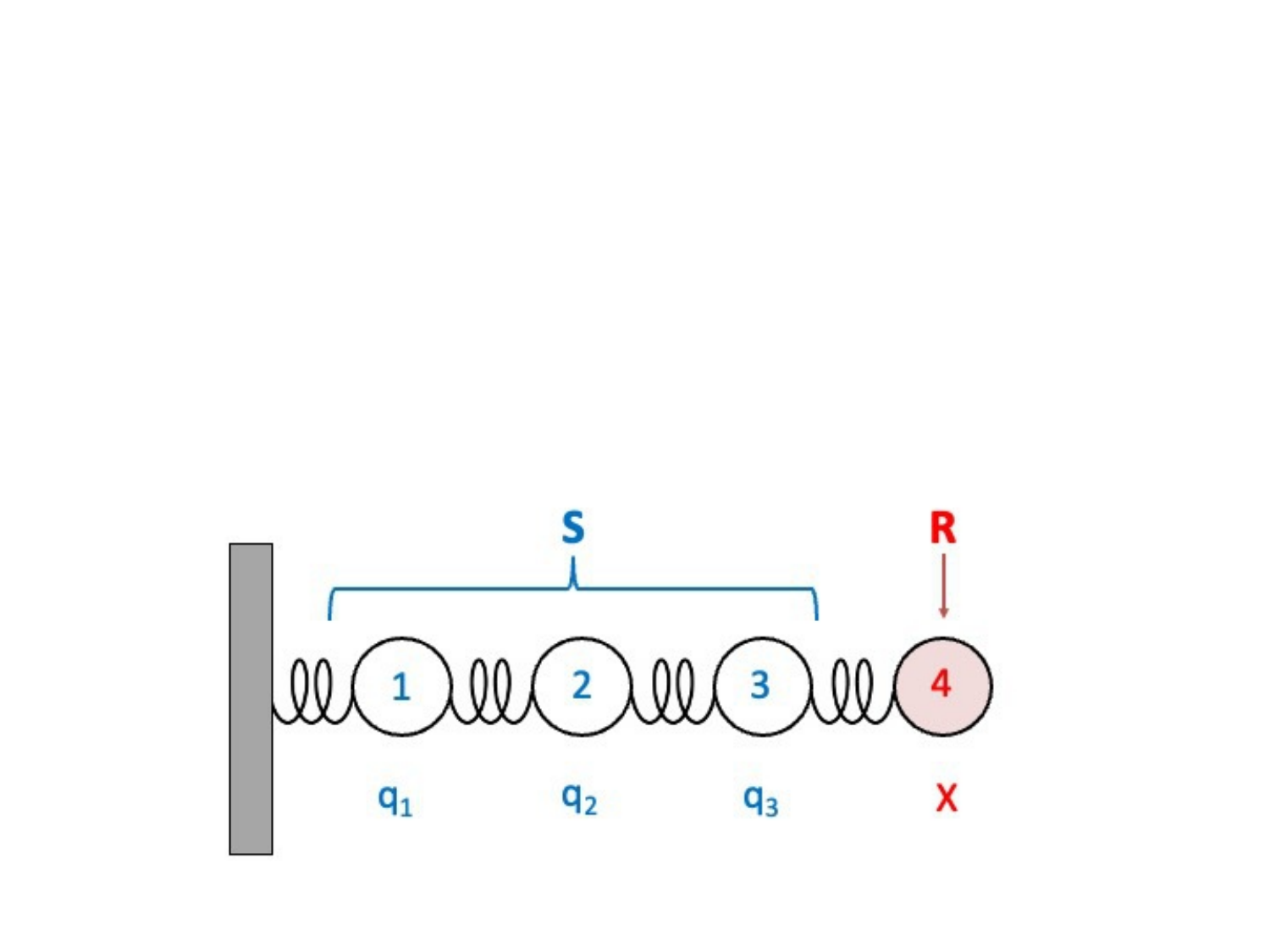}
\caption{
Toy model illustrating our general setup.
${\cal S}$ consists of three particles of mass $m$, and ${\cal R}$ is a particle of mass $M$.
The wall at the left is fixed, and springs denote harmonic couplings.
}
\label{fig:toyModel}
\end{figure}

\section*{The system and work source are thermally isolated}

Here we obtain Eq.~\ref{eq:ft-a-intro} for a system and work source evolving in the absence of a thermal reservoir.
To begin, we specify our setup and define the quantities $W$, $\Delta F$ and $\Delta\phi$.

Consider a classical system, ${\cal S}$, with $D$ degrees of freedom.
Let ${\bf z} = ({\bf q},{\bf p})$ denote a point in ${\cal S}$'s phase space, where ${\bf q}$ and ${\bf p}$ denote coordinates and momenta.
${\cal S}$ interacts with a body, or work source, ${\cal R}$, with one degree of freedom and mass $M$.
Let $(X,P)$ denote a point in ${\cal R}$'s phase space.
The combined system and work source, ${\cal RS}$, is described by a Hamiltonian
\be
\label{eq:HRS}
H_{\cal RS}(X,P,{\bf q},{\bf p}) = \frac{P^2}{2M} + H_{\cal S}({\bf z};X) \equiv H_{\cal R}^0 + H_{\cal S}
\, .
\ee
Since there is no thermal reservoir (for now), ${\cal RS}$ evolves under Hamiltonian dynamics, generated by $H_{\cal RS}$.

Throughout most of this paper, we view the kinetic term $H_{\cal R}^0=P^2/2M$ as the energy of the work source, and $H_{\cal S}({\bf z};X)$ as the energy of the system of interest, parametrized by the coordinate $X$.
This particular partitioning of the total energy, $H_{\cal RS}$, between subsystems ${\cal R}$ and ${\cal S}$ corresponds to the {\it inclusive} convention~\cite{Jarzynski2007}.
In this convention, all interactions between ${\cal S}$ and ${\cal R}$ are counted as contributing to the energy of ${\cal S}$.
However, in one section below, see Eqs.~\ref{eq:HS-split}-\ref{eq:ft-na-exc}, we will temporarily adopt the {\it exclusive} convention, in which ${\cal S}-{\cal R}$ interactions are treated as contributing to ${\cal R}$'s energy.
Both conventions are thermodynamically valid, and they lead to different fluctuation theorems for non-autonomous work~\cite{Jarzynski2007}.
We will similarly find different fluctuation theorems for inclusive and exclusive autonomous work (Eqs.~\ref{eq:ft-a-intro}, \ref{eq:ft-exc}).

Figure~\ref{fig:toyModel} illustrates our setup.
Here we have
\be
\label{eq:HS}
H_{\cal S}({\bf q},{\bf p};X) = \sum_{n=1}^3 \left[ \frac{p_n^2}{2m} + \frac{k}{2} (q_n - q_{n-1}-l)^2 \right] + \frac{k}{2} (X - q_3-l)^2 \,,
\ee
where $k,\, l > 0$, and $q_0\equiv 0$ is the location of a fixed wall.
$H_{\cal S}$ includes the kinetic energies of the three particles that compose ${\cal S}$, and the potential energies of all four springs.
Adding the kinetic energy $P^2/2M$ to $H_{\cal S}$, we obtain the Hamiltonian $H_{\cal RS}$ that governs the evolution of the four-particle collection ${\cal RS}$.
While this toy model is useful to keep in mind, the following derivation applies generally to a system and work reservoir described by the Hamiltonian $H_{\cal RS}$ given by Eq.~\ref{eq:HRS}.

It will prove convenient to work with ${\cal R}$'s velocity, $V=P/M$, rather than its momentum.
Let $\Gamma = (X,V,{\bf q},{\bf p})$ denote a microscopic state, or {\it microstate}, of ${\cal RS}$.
Hamilton's equations of motion read,
\be
\label{eq:hameq}
\dot X = V \quad,\quad \dot V = -\frac{1}{M} \frac{\partial H_{\cal S}}{\partial X}  \quad,\quad
\dot{\bf q} = \frac{\partial H_{\cal S}}{\partial{\bf p}} \quad,\quad \dot{\bf p} = -\frac{\partial H_{\cal S}}{\partial{\bf q}} \, .
\ee
Throughout the paper, dots denote total time derivatives.
Under these dynamics, $H_{\cal R}^0$ and $H_{\cal S}$ vary with time, but their sum, $H_{\cal RS}$, remains constant.
That is, ${\cal R}$ and ${\cal S}$ exchange energy, but the total energy is fixed.

Autonomous work is the transfer of energy from the work source to the system of interest.
Therefore, ${\cal R}$ performs work on ${\cal S}$ at a rate $\dot W = -\dot H_{\cal R}^0 = \dot H_{\cal S}$.
Integrating these expressions over a finite time interval $[0,\tau]$, we obtain three equivalent expressions for the work performed on ${\cal S}$ during this interval:
\ba
\label{eq:W}
W[\Gamma_t] &=& \frac{M}{2} V_0^2 - \frac{M}{2} V_\tau^2
= H_{\cal S}({\bf z}_\tau;X_\tau) - H_{\cal S}({\bf z}_0;X_0) \nonumber \\
&=& \int_{0}^{\tau} dt \, \dot X_t \,  \frac{\partial H_{\cal S}}{\partial X}({\bf z}_t;X_t) \, .
\ea
The first two expressions for $W$ represent the energy lost by ${\cal R}$ and that gained by ${\cal S}$, while the third is an integrated displacement ($dX$) $\times$ force ($\partial H_{\cal S}/\partial X$).
The final equality in Eq.~\ref{eq:W} follows from Eq.~\ref{eq:hameq}.
The notation $W[\Gamma_t]$ indicates that work is a functional of the trajectory $\Gamma_t$ that describes the evolution of ${\cal RS}$ from $t=0$ to $\tau$.

For any coordinate value $X$ and temperature $T=\beta^{-1}$, let
\be
\label{eq:pi}
\pi({\bf z}\vert X) = e^{\beta [F(X) - H_{\cal S}({\bf z};X)]}
\ee
denote a normalized equilibrium distribution of ${\cal S}$, conditioned on ${\cal R}$'s coordinate.
We refer to $\pi({\bf z}\vert X)$ as a {\it conditional equilibrium state} of ${\cal S}$ \footnote{This conditional state is similar in spirit to the conditional stochastic work introduced in Ref.~\cite{Sone2021_JSP}.  The present scenario is, however, more general, as we are not restricted to a specific choice of $X$.
In the setting of strong-coupling thermodynamics, Refs.~\cite{Seifert2016}, \cite{Talkner2016} and \cite{Miller2017} use similarly defined equilibrium states of a heat bath, ${\cal T}$, conditioned on microstates of ${\cal S}$.}.
We can imagine preparing this state by allowing the system of interest to equilibrate with a thermal reservoir, with $X$ held fixed.
$F(X)$ is the free energy of ${\cal S}$ in the state $\pi({\bf z}\vert X)$.
Throughout this paper, the dependence of both $\pi$ and $F$ on temperature is left implicit.

Imagine a process in which we prepare ${\cal RS}$ in an initial statistical state
\be
\label{eq:f0}
f(\Gamma,0) = \rho_0(X,V) \, \pi({\bf z}\vert X) \,,
\ee
where $\rho_0$ is an arbitrary distribution on ${\cal R}$'s phase space.
For example, prior to $t=0$ we might prepare ${\cal RS}$ in an equilibrium state, with ${\cal R}$'s coordinate $X$ subject to a quadratic confining potential $(\kappa/2)(X-\bar X)^2$.
For sufficiently large $\kappa$, this amounts to setting ${\cal R}$'s initial value at approximately $\bar X$.
At $t=0$ we release ${\cal R}$ by turning off the confining potential.
Additionally, we might subject ${\cal R}$ to an impulsive force $M\Delta V \delta(t-0^-)$, imparting a sudden change $\Delta V$ to its initial velocity.
More intricate methods of preparing ${\cal R}$'s state are also possible.
However ${\cal R}$ is prepared, we assume ${\cal S}$ begins in the conditional equilibrium state $\pi({\bf z}\vert X)$.
This assumption mirrors the assumption of an initial equilibrium state that is made when deriving Eq.~\ref{eq:ft-na}~\cite{Jarzynski1997a,Jarzynski1997b}.

For one realization of the process, a microstate $\Gamma_0$ is drawn randomly from $f(\Gamma,0)$.
Then, from $t=0$ to $\tau$, ${\cal RS}$ evolves under Eq.~\ref{eq:hameq}, from the initial microstate $\Gamma_0$.
Repeating this process, we obtain an ensemble of realizations.
Each realization is described by trajectory $\Gamma_t$, and the ensemble is described by a probability distribution $f(\Gamma,t)$ that evolves under Liouville's equation,
\be
\label{eq:liouville}
\frac{\partial f}{\partial t} = -V \frac{\partial f}{\partial X} + \frac{1}{M}\frac{\partial H_{\cal S}}{\partial X} \frac{\partial f}{\partial V} 
- \frac{H_{\cal S}}{\partial {\bf p}} \cdot \frac{\partial f}{\partial {\bf q}} + \frac{H_{\cal S}}{\partial {\bf q}} \cdot \frac{\partial f}{\partial {\bf p}}
\, .
\ee
By Liouville's theorem~\cite{Goldstein1980}, $(d/dt) f(\Gamma_t,t) = 0$ along any trajectory $\Gamma_t$ in the ensemble.
Let
\be
\label{eq:rho}
\rho_t(X,V) = \int d{\bf z} \, f(X,V,{\bf z},t)
\ee
denote the evolving marginal distribution of ${\cal R}$.

For a single trajectory $\Gamma_t$, we define
\ba
\label{eq:DeltaF}
\Delta F[\Gamma_t] &=& F(X_\tau) - F(X_0) \\
\label{eq:Deltaphi}
\Delta\phi[\Gamma_t] &=& -\ln\rho_\tau(X_\tau,V_\tau) + \ln\rho_0(X_0,V_0) \,.
\ea
We view $\Delta F$ as a change in the free energy change of ${\cal S}$, and $\Delta\phi$ as a change in the entropy of ${\cal R}$, along a particular trajectory.
The values of $W$, $\Delta F$ and $\Delta\phi$ fluctuate from one realization of the process to another. $\Delta\phi$ is a standard definition of stochastic entropy production~\cite{Seifert2012,Peliti2021,Limmer2024,Seifert2025}. In the non-autonomous setting, $\Delta F$ is a change in a state function and, hence, does not fluctuate. Here, however, $\Delta F$ is a stochastic free energy change, whose fluctuations are driven by the dynamics of the work source.
Angular brackets, as in Eq.~\ref{eq:ft-a-intro}, denote an average over the ensemble of realizations.

We now have the elements in place to derive Eq.~\ref{eq:ft-a-intro}.
Since $\Gamma_t$ evolves deterministically, $W[\Gamma_t]$, $\Delta F[\Gamma_t]$ and $\Delta\phi[\Gamma_t]$ are determined by the initial conditions $\Gamma_0$.
We thus have
\begin{align}
\label{eq:ft}
&\left\langle e^{-\beta ( W - \Delta F ) - \Delta\phi} \right\rangle
= \int d\Gamma_0 \, f(\Gamma_0,0) \, e^{-\beta ( W - \Delta F ) - \Delta\phi} \nonumber\\
&=  \int d\Gamma_0 \,  \rho_0(X_0,V_0) \, \pi({\bf z}_0 \vert X_0) \, \frac{ \pi({\bf z}_\tau \vert X_\tau )}{ \pi({\bf z}_0 \vert X_0) } \, \frac{ \rho_\tau(X_\tau,V_\tau) }{ \rho_0(X_0,V_0) } \nonumber\\
&= \int d\Gamma_\tau \,  \rho_\tau(X_\tau,V_\tau) \, \pi({\bf z}_\tau \vert X_\tau) = 1 \, .
\end{align}
In going from the first to the second line, we use Eqs.~\ref{eq:W}, \ref{eq:pi}, \ref{eq:f0}, \ref{eq:DeltaF} and \ref{eq:Deltaphi}.
Since $\vert \partial\Gamma_\tau/\partial\Gamma_0\vert=1$ (by Liouville's theorem), we replace integration over initial conditions with integration over final conditions to get to the third line.

Equation~\ref{eq:ft} is the autonomous counterpart of Eq.~\ref{eq:ft-na}.
By Jensen's inequality, Eq.~\ref{eq:ft} implies
\be
\label{eq:2lt}
\beta \langle W - \Delta F \rangle + \Delta S_{\cal R} \ge 0 \, ,
\ee
where
\be
\Delta S_{\cal R} \equiv \langle\Delta\phi\rangle = -\int \rho_\tau\ln\rho_\tau + \int \rho_0\ln\rho_0 \, .
\ee
Equation~\ref{eq:2lt} is a second-law inequality.
The term $\Delta S_{\cal R}$ is the change in the Shannon entropy of the work source.
Since ${\cal R}$ is described by a single coordinate, its Shannon entropy is entirely negligible when ${\cal S}$ is macroscopic; in that setting Eq.~\ref{eq:2lt} reduces to the classic statement of the second law given by Eq.~\ref{eq:2lt-na}.
When ${\cal S}$ is microscopic, $\Delta S_{\cal R}$ is not necessarily negligible.

\section*{The system and work source interact with a heat bath}

To this point, we have assumed ${\cal S}$ and ${\cal R}$ exchange energy with one another, but not with other bodies.
Now imagine that ${\cal S}$ additionally exchanges energy with a thermal reservoir, or heat bath, ${\cal T}$, at temperature $T=\beta^{-1}$, as illustrated in Fig.~\ref{fig:global-1d}a.
To derive Eq.~\ref{eq:ft-a-intro} in this setting, we can proceed by introducing the bath's degrees of freedom explicitly, 
then using Hamiltonian dynamics to model the evolution of the combined system of interest, work source and heat bath, ${\cal RST}$.
Alternatively, we can treat the heat bath implicitly, using stochastic dynamics to model its effects on ${\cal RS}$.
The implicit treatment is simpler, and we take that route here, leaving the explicit treatment to the Supporting Information.

We begin by assuming that if ${\cal R}$ is fixed at a coordinate value $X$, then ${\cal S}$ evolves under stochastic, Markovian dynamics.
We represent these dynamics by equations of motion that depend parametrically on $X$:
\be
\label{eq:stochEOM-z}
\dot{\bf z} = {\bm\psi}({\bf z};X) \, .
\ee
(See Eq.~\ref{eq:lang} for an example.)
An ensemble of trajectories, evolving independently under Eq.~\ref{eq:stochEOM-z}, is described by a probability density $\varphi({\bf z},t)$ that obeys
\be
\label{eq:me-z}
\frac{\partial\varphi}{\partial t} = \hat{\cal L}_{\cal S}(X) \varphi \, .
\ee
The linear operator $\hat{\cal L}_{\cal S}(X)$ acts on the variables ${\bf z}$, and depends parametrically on $X$ (see e.g.\ Eq.~\ref{eq:LS-example}).
We assume this operator satisfies
\be
\label{eq:wdb-z}
\hat{\cal L}_{\cal S}(X) \pi({\bf z} \vert X) = 0 \, .
\ee
That is, the stochastic dynamics $\dot{\bf z} = {\bm\psi}({\bf z};X)$ preserve the distribution $\pi({\bf z} \vert X)$.
This assumption is physically motivated: it states that if ${\cal S}$ begins in thermal equilibrium and $X$ is held fixed, then interactions with ${\cal T}$ do not drive ${\cal S}$ away from equilibrium.

When both ${\cal R}$ and ${\cal S}$ evolve with time, their joint evolution obeys the following dynamics:
\be
\label{eq:stochEOM}
\dot X = V \quad,\quad \dot V = -\frac{1}{M} \frac{\partial H_{\cal S}}{\partial X}({\bf z};X)  \quad,\quad
\dot{\bf z} = {\bm\psi}({\bf z};X) \, .
\ee
At the ensemble level of description, the probability density $f(\Gamma,t)$ satisfies
\be
\label{eq:me}
\frac{\partial f}{\partial t} = -V \frac{\partial f}{\partial X} + \frac{1}{M}\frac{\partial H_{\cal S}}{\partial X} \frac{\partial f}{\partial V} + \hat{\cal L}_{\cal S}(X) f \equiv \hat{\cal L}_{\cal RS} f \, .
\ee
Equations~\ref{eq:HRS}, \ref{eq:pi} and \ref{eq:wdb-z} imply
\be
\label{eq:wdb-Gamma}
\hat{\cal L}_{\cal RS} \, e^{-\beta H_{\cal RS}} = 0 \, .
\ee
The equilibrium-preserving property of $\hat{\cal L}_{\cal S}(X)$ (Eq.~\ref{eq:wdb-z}) is promoted to the full phase space when both ${\cal S}$ and ${\cal R}$ evolve (Eq.~\ref{eq:wdb-Gamma}).

As always, a concrete example is helpful.
For the toy model shown in Fig.~\ref{fig:toyModel}, the equations of motion
\be
\label{eq:lang}
\dot{\bf q} = \frac{\partial H_{\cal S}}{\partial{\bf p}} \quad,\quad
\dot{\bf p} = -\frac{\partial H_{\cal S}}{\partial{\bf q}} - \gamma\frac{\bf p}{m}  + \sqrt{2\gamma T} \, \bm{\xi}(t)
\ee
offer a natural choice for the dynamics $\dot{\bf z} = {\bm\psi}({\bf z};X)$.
The dependence of ${\bm\psi}$ on $X$ enters through $H_{\cal S}$ (Eq.~\ref{eq:HS}).
$\gamma$ is a friction coefficient, and $\bm{\xi}(t)$ is a noise vector whose components satisfy $\langle \xi_i(s)\xi_j(t)\rangle =  \delta_{ij}\delta(t-s)$.
In Eq.~\ref{eq:lang}, Brownian dynamics model the evolution of ${\cal S}$.
The operator $\hat{\cal L}_{\cal S}(X)$ describing this evolution at the ensemble level is given by:
\be
\label{eq:LS-example}
\hat{\cal L}_{\cal S}(X) \varphi =
-\frac{\partial H_{\cal S}}{\partial{\bf p}}\cdot\frac{\partial\varphi}{\partial{\bf q}}
+\frac{\partial H_{\cal S}}{\partial{\bf q}}\cdot\frac{\partial\varphi}{\partial{\bf p}}
+ \frac{\gamma}{m} \frac{\partial}{\partial{\bf p}}\cdot({\bf p} \varphi) + \gamma T \frac{\partial}{\partial{\bf p}} \cdot \frac{\partial \varphi}{\partial{\bf p}} \, .
\ee
The terms on the right describe Hamiltonian evolution, friction and momentum diffusion.
The joint evolution of ${\cal RS}$ is governed by the equation $\partial f/\partial t = \hat{\cal L}_{\cal RS} f$, with $\hat{\cal L}_{\cal RS}$ given by Eq.~\ref{eq:me}.
By inspection, Eqs.~\ref{eq:wdb-z} and \ref{eq:wdb-Gamma} are satisfied.
While the toy model serves as a useful example, the following derivation is valid more generally, for dynamics described by Eqs.~\ref{eq:stochEOM} - \ref{eq:wdb-Gamma}.

The propagator
\be
\label{eq:propagator}
K_t(\Gamma \vert \Gamma_0) = \exp\left( \hat{\cal L}_{\cal RS} \, t \right)  \delta\left( \Gamma - \Gamma_0 \right)
\ee
gives the probability density to find ${\cal RS}$ in the microstate $\Gamma$ at time $t_0+t$, given that it was in the microstate $\Gamma_0$ at an earlier time $t_0$.
Because Eq.~\ref{eq:stochEOM} describes a stationary process, $K_t$ does not depend on $t_0$.
Equation~\ref{eq:wdb-Gamma} implies, for all $t>0$,
\be
\label{eq:wdbK}
\int d\Gamma_0 \, K_t(\Gamma \vert \Gamma_0) \, e^{ -\beta H_{\cal RS}(\Gamma_0) } =e^{ -\beta H_{\cal RS}(\Gamma) } \, .
\ee

We now derive Eq.~\ref{eq:ft-a-intro}.
Imagine a process in which initial conditions for ${\cal RS}$ are sampled from $f(\Gamma,0)=\rho_0\pi$ (Eq.~\ref{eq:f0}).
Then, from $t=0$ to $\tau$, ${\cal RS}$ evolves under Eq.~\ref{eq:stochEOM}.
Every realization produces a trajectory $\Gamma_t$, and the ensemble of trajectories is described by an evolving distribution,
\be
f(\Gamma,t) = \int d\Gamma_0 \, K_t(\Gamma \vert \Gamma_0) \, f(\Gamma_0,0) \, .
\ee
For each realization of the process, the work performed on ${\cal S}$ is the energy lost by ${\cal R}$:
\begin{align}
W[\Gamma_t] &=  \frac{M}{2} V_0^2 - \frac{M}{2} V_\tau^2 \nonumber \\
\label{eq:Wstoch}
&= H_{\cal S}({\bf z}_\tau;X_\tau) - H_{\cal S}({\bf z}_0;X_0) - H_{\cal RS}(\Gamma_\tau) + H_{\cal RS}(\Gamma_0) \, ,
\end{align}
where the second line follows from Eq.~\ref{eq:HRS}.
We again define $\Delta F[\Gamma_t]$ and $\Delta\phi[\Gamma_t]$ by Eqs.~\ref{eq:DeltaF} and \ref{eq:Deltaphi}.
Since $W$, $\Delta F$ and $\Delta\phi$ are determined by the initial and final microstates, $\Gamma_0$ and $\Gamma_\tau$, we derive Eq.~\ref{eq:ft-a-intro} by integrating over these microstates:
\begin{align}
\label{eq:ft-a-rst}
&\left\langle e^{-\beta ( W - \Delta F ) - \Delta\phi} \right\rangle
= \int d\Gamma_0 \int d\Gamma_\tau f(\Gamma_0,0) K_\tau e^{-\beta(W-\Delta F) - \Delta\phi} \nonumber \\
&= \int d\Gamma_0 \int d\Gamma_\tau \rho_\tau(X_\tau,V_\tau) \pi({\bf z}_\tau\vert X_\tau) K_\tau e^{\beta H_{\cal RS}(\Gamma_\tau)} e^{- \beta H_{\cal RS}(\Gamma_0) } \nonumber \\
&= \int d\Gamma_\tau \, \rho_\tau(X_\tau,V_\tau) \, \pi({\bf z}_\tau\vert X_\tau)  = 1 \, ,
\end{align}
where $K_\tau=K_\tau(\Gamma_\tau \vert \Gamma_0)$.
On the first line, the product $fK_\tau$ is the joint probability density for the initial and final microstates $\Gamma_0$ and $\Gamma_\tau$.
We use Eqs.~\ref{eq:pi}, \ref{eq:f0}, \ref{eq:DeltaF}, \ref{eq:Deltaphi}, and \ref{eq:Wstoch} to get to the second line, and Eq.~\ref{eq:wdbK} to get to the third.

We conclude that Eq.~\ref{eq:ft-a-intro} is valid when ${\cal RS}$ evolves under the Markovian dynamics given by Eq.~\ref{eq:stochEOM}.
Since the Hamiltonian dynamics of Eq.~\ref{eq:hameq} are a special case of Eq.~\ref{eq:stochEOM}, for the rest of this paper we simply assume that ${\cal RS}$ evolves under Eq.~\ref{eq:stochEOM} at the single-trajectory level, and Eq.~\ref{eq:me} at the ensemble level.

\section*{Recovering Eq.~\ref{eq:ft-na} when $M\rightarrow\infty$}

To relate Eq.~\ref{eq:ft-a-intro} to Eq.~\ref{eq:ft-na}, consider what happens when ${\cal R}$ has infinite inertia.
In this limit, general arguments suggest that the autonomous definition of work reduces to the non-autonomous one~\cite{Jarzynski1998}.
For a single trajectory $\Gamma_t$ evolving from initial conditions $\Gamma_0$, the limit $M\rightarrow\infty$ combines with either Eq.~\ref{eq:hameq} or \ref{eq:stochEOM} to give
\be
\label{eq:Xevoln-na}
X_t = X_0 + V_0 t \quad,\quad V_t = V_0 \quad,\quad 0\le t\le\tau \, .
\ee
The infinitely massive external body moves at constant velocity $V_0$, experiencing no back-action, and the system of interest evolves under the Hamiltonian $H_{\cal S}({\bf z};X_0 + V_0 t)$.
For an ensemble of realizations, Eq.~\ref{eq:Xevoln-na} implies
\be
\label{eq:rhoShear}
\rho_\tau(X,V) = \rho_0(X-V\tau,V) \, ,
\ee
which in turn gives
\be
\label{eq:dphi0}
\Delta\phi[\Gamma_t] = 0 \, .
\ee
In this situation there is no change in ${\cal R}$'s Shannon entropy, $\Delta S_{\cal R}=0$, and Eqs.~\ref{eq:ft} and \ref{eq:2lt} become
\be
\label{eq:infiniteInertia}
\left\langle e^{-\beta ( W - \Delta F )} \right\rangle = 1 \quad,\quad \langle W - \Delta F \rangle \ge 0 \, .
\ee

Thus when $M\rightarrow\infty$, back-action no longer torments ${\cal R}$'s evolution, and the terms $\Delta\phi$ and $\Delta S_{\cal R}$ vanish from Eqs.~\ref{eq:ft-a-intro} and \ref{eq:2lt-a-intro}.
However, ${\cal R}$'s evolution still depends on its initial conditions $(X_0,V_0)$ (Eq.~\ref{eq:Xevoln-na}), hence so does $\Delta F$ (Eq.~\ref{eq:DeltaF}).
Let us therefore additionally suppose that these initial conditions are $\delta$-distributed:
\be
\label{eq:delta}
\rho_0(X,V) = \delta(X-X_A) \, \delta(V-V_A) \, ,
\ee
for a particular choice $(X_A,V_A)$.
We then obtain
\be
\label{eq:dF-noFluctuations}
\Delta F[\Gamma_t] = F(X_B) - F(X_A) \quad,\quad X_B = X_A + V_A\tau
\ee
for every trajectory $\Gamma_t$.
Now $\Delta F$ no longer fluctuates from one realization to the next, and Eq.~\ref{eq:infiniteInertia} becomes:
\be
\label{eq:ft-a-infinite+delta}
\left\langle e^{-\beta W} \right\rangle = e^{-\beta \Delta F }
\quad,\quad
\langle W \rangle \ge  \Delta F \, .
\ee

The equivalence between Eq.~\ref{eq:ft-na} and the equality in Eq.~\ref{eq:ft-a-infinite+delta} has a clear interpretation.
In the non-autonomous case, for every realization of the process, the work parameter follows the same externally imposed protocol, $\lambda_t$, hence the system of interest evolves under the same time-dependent Hamiltonian, $H_{\cal S}({\bf z};\lambda_t)$.
In the autonomous case, with $M\rightarrow\infty$ and Eq.~\ref{eq:delta}, the work source evolves along the same trajectory $(X_t,V_t)$ during every realization, and the system evolves under the same time-dependent equation, $\dot{\bf z} = {\bm\psi}({\bf z};X_t)$.
${\cal S}$ has no way of ``knowing'' whether the time-dependence of $H_{\cal S}$ is due to an externally manipulated work parameter, or an infinitely massive work source.

For an autonomous work source with infinite inertia, the inequality in Eq.~\ref{eq:ft-a-infinite+delta} was derived in Ref.~\cite{Deffner2013_prx}, in a generalized form (see Eq.~58 therein), within a Hamiltonian framework.

\section*{Fluctuation theorem for exclusive work}

Let us split $H_{\cal S}$ into a  system self-energy, $H_{\cal S}^0$, and a system-reservoir interaction energy, $H_{\rm int}$:
\be
\label{eq:HS-split}
H_{\cal S}({\bf z};X) = H_{\cal S}^0({\bf z}) + H_{\rm int}({\bf z},X) \, .
\ee
This separation, while not mathematically unique, is often physically motivated.
For the toy model of Fig.~\ref{fig:toyModel}, it is natural to identify $H_{\rm int}$ with the potential energy stored in the rightmost spring, as already apparent in Eq.~\ref{eq:HS}.
The total Hamiltonian is
\be
\label{eq:HRS-exclusive}
H_{\cal RS}(\Gamma) = H_{\cal R}^0 + H_{\cal S}^0 + H_{\rm int} \, ,
\ee
with $H_{\cal R}^0 = MV^2/2$, as before.

In earlier sections, we defined the energies of the work source and system of interest to be $H_{\cal R}^0$ and $H_{\cal S}$, respectively. Now suppose we instead define the energies of ${\cal R}$ and ${\cal S}$ to be $H_{\cal R}\equiv H_{\cal R}^0+H_{\rm int}$ and $H_{\cal S}^0$.
We now have $H_{\cal RS} = H_{\cal R} + H_{\cal S}^0$, and the work performed on ${\cal S}$ (the energy lost by ${\cal R}$) is
\begin{align}
W^0&[\Gamma_t] = H_{\cal R}(X_0,V_0;{\bf z}_0) - H_{\cal R}(X_\tau,V_\tau;{\bf z}_\tau) \nonumber \\
\label{eq:W0}
&= H_{\cal RS}(\Gamma_0) - H_{\cal RS}(\Gamma_\tau) + H_S^0({\bf z}_\tau) - H_S^0({\bf z}_0) \, .
\end{align}
We can use Eq.~\ref{eq:stochEOM} to write $W^0$ as an integrated displacement $\times$ force,
\be
\label{eq:W0fdx}
W^0[\Gamma_t] = - \int_0^\tau dt \, \dot{\bf z}_t \cdot \frac{\partial H_{\cal R}}{\partial{\bf z}}(X_t,V_t;{\bf z}_t) \, ,
\ee
which is similar in form to the final expression in Eq.~\ref{eq:W}.\footnote{
Both $H_{\cal S}$ in Eq.~\ref{eq:W} and $H_{\cal R}$ in Eq.~\ref{eq:W0fdx} can be replaced by $H_{\rm int}$, highlighting that the difference $W-W^0$ is the net change in the energy stored in the interaction term.}

The difference between $W$ (Eqs.~\ref{eq:W}, \ref{eq:Wstoch}) and $W^0$ (Eq.~\ref{eq:W0}) resides in whether we interpret the interaction energy $H_{\rm int}$ as belonging to ${\cal S}$ or ${\cal R}$.
We will refer to $W$ and $W^0$, respectively, as {\it inclusive} and {\it exclusive} autonomous work, following terminology used for non-autonomous work~\cite{Jarzynski2007}.
We now derive a fluctuation theorem for $W^0$.

Suppose we sample initial conditions for ${\cal RS}$ from the distribution
\be
\label{eq:f00}
f(\Gamma,0) = \rho_0(X,V) \, \pi^0({\bf z}) \, ,
\ee
where $\rho_0$ is arbitrary (as earlier), and
\be
\label{eq:pi0}
\pi^0({\bf z}) = e^{\beta[F^0 - H_{\cal S}^0({\bf z})]}
\ee
is the canonical distribution for the Hamiltonian $H_{\cal S}^0({\bf z})$.
We then obtain
\begin{align}
\label{eq:ft-exc}
&\left\langle e^{-\beta W^0 - \Delta\phi} \right\rangle
= \int d\Gamma_0 \int d\Gamma_\tau \, f(\Gamma_0,0) \, K_\tau \, e^{-\beta W^0 - \Delta\phi} \nonumber \\
&= \int d\Gamma_0 \int d\Gamma_\tau \,  \rho_\tau(X_\tau,V_\tau) \, \pi^0({\bf z}_\tau) \, K_\tau \, e^{\beta H_{\cal RS}(\Gamma_\tau)} e^{- \beta H_{\cal RS}(\Gamma_0) } \nonumber \\
&= \int d\Gamma_\tau \, \rho_\tau(X_\tau,V_\tau) \, \pi^0({\bf z}_\tau)  = 1 \, ,
\end{align}
where $K_\tau=K_\tau(\Gamma_\tau \vert \Gamma_0)$.
We used Eqs.~\ref{eq:Deltaphi}, \ref{eq:W0}, \ref{eq:f00} and \ref{eq:pi0} to get to the second line, and Eq.~\ref{eq:wdbK} to get to the third.
We conclude that both inclusive and exclusive definitions of autonomous work support fluctuation theorems (Eq.~\ref{eq:ft-a-intro}, \ref{eq:ft-exc}).
In each case initial conditions are sampled canonically (Eqs.~\ref{eq:f0}, \ref{eq:f00}), according to the Hamiltonian that represents the system's internal energy ($H_{\cal S}$, $H_{\cal S}^0$).

Equation~\ref{eq:ft-exc} is the autonomous counterpart of the non-autonomous fluctuation theorem for exclusive work,
\begin{equation}
\label{eq:ft-na-exc}
\langle e^{-\beta W^0}\rangle = 1 \, ,
\end{equation}
due to Bochkov and Kuzovlev~\cite{Bochkov1977,Bochkov1981}.
Eq.~\ref{eq:ft-exc} reduces to Eq.~\ref{eq:ft-na-exc} when $M\rightarrow\infty$ and ${\cal R}$'s initial conditions are delta-distributed (Eq.~\ref{eq:delta}), as was the case with Eqs.~\ref{eq:ft-a-intro} and \ref{eq:ft-na}.

We end this section with brief comments on the preparation of the state $f=\rho_0\pi^0$ (Eq.~\ref{eq:f00}).
The initial equilibrium state $\pi^0({\bf z})$ can be motivated by imagining that, prior to $t=0$, the system of interest equilibrates with a heat bath, with the ${\cal S}-{\cal R}$ interaction term, $H_{\rm int}({\bf z},X)$, turned off.
That is, ${\cal S}$ and ${\cal R}$ are initially decoupled.
At $t=0$, $H_{\rm int}$ is turned on: ${\cal S}$ and ${\cal R}$ are coupled.
Subsequently ${\cal SR}$ evolves under $H_{\cal RS}$ (Eq.~\ref{eq:HRS-exclusive}).
Of course, the idea that $H_{\rm int}$ can be turned on and off like a light switch is an idealization.
Its plausibility depends on details of these subsystems.
Additionally, turning on the interaction term requires work, which is performed by the entity that prepares the initial state.
Since $H_{\rm int}$ is not counted as contributing to the energy of ${\cal S}$, in the exclusive convention, this work is performed on the work source itself, and not on the system of interest.

\section*{Fluctuation theorem for total entropy production}

Let us return to the inclusive definition of autonomous work, Eq.~\ref{eq:Wstoch}.
By the first law of thermodynamics, the heat absorbed by ${\cal S}$ is
\be
\label{eq:Q}
Q[\Gamma_t] = \Delta H_{\cal S} - W = H_{\cal RS}(\Gamma_\tau) - H_{\cal RS}(\Gamma_0) \, .
\ee
Suppose initial conditions for ${\cal RS}$ are sampled from an arbitrary distribution $f(\Gamma,0)$, not necessarily satisfying Eq.~\ref{eq:f0}.
For $t\ge 0$, $f(\Gamma,t)$ evolves under $\partial f/\partial t= \hat{\cal L}_{\cal RS} f$ (Eq.~\ref{eq:me}).
Without loss of generality, we write
\be
\label{eq:fGt}
f(\Gamma,t) = \rho_t(X,V) \, \eta_t({\bf z} \vert X,V) \, .
\ee
Here, $\rho_t = \int d{\bf z} \, f$ (Eq.~\ref{eq:rho}) is ${\cal R}$'s distribution at time $t$; and $\eta_t$ is ${\cal S}$'s distribution, conditioned on ${\cal R}$'s microstate.

Along a trajectory $\Gamma_t$, $-\ln\eta_\tau({\bf z}_t \vert X_t,V_t)$ is the system of interest's conditional stochastic entropy at time $t$.
Let
\be
\label{eq:Deltasigma}
\Delta\sigma[\Gamma_t] \equiv -\ln\eta_\tau({\bf z}_\tau \vert X_\tau,V_\tau) + \ln\eta_0({\bf z}_0 \vert X_0,V_0) \, ,
\ee
denote the net change in this conditional stochastic entropy, over one realization of the process.
Finally, let
\be
\Delta s_{\rm tot} \equiv -\beta Q +  \Delta\sigma + \Delta\phi
\ee
denote a total entropy change, in the sense that the terms on the right represent entropy changes in ${\cal T}$, ${\cal S}$ and ${\cal R}$, respectively.
We have, with $K_\tau=K_\tau(\Gamma_\tau \vert \Gamma_0)$,
\begin{align}
\left\langle e^{-\Delta s_{\rm tot}} \right\rangle
&= \int d\Gamma_0 \int d\Gamma_\tau \, f(\Gamma_0,0) \, K_\tau \, e^{\beta Q - \Delta\sigma - \Delta\phi} \nonumber \\
&= \int d\Gamma_0 \int d\Gamma_\tau \,  f(\Gamma_\tau,\tau) \, K_\tau \, e^{\beta H_{\cal RS}(\Gamma_\tau)} e^{- \beta H_{\cal RS}(\Gamma_0) } \nonumber \\
\label{eq:sft-a}
&= \int d\Gamma_\tau \, f(\Gamma_\tau,\tau)  = 1 \, ,
\end{align}
using Eqs.~\ref{eq:Deltaphi}, \ref{eq:wdbK}, \ref{eq:Q}, \ref{eq:fGt} and \ref{eq:Deltasigma}.

Equation~\ref{eq:sft-a} is the autonomous version of an integral fluctuation theorem for entropy production~\cite{Crooks1999,Jarzynski1999a,Seifert2005}.
Equation~\ref{eq:sft-a} reduces to its non-autonomous counterpart when $M\rightarrow\infty$ and ${\cal R}$'s initial conditions are delta-distributed (Eq.~\ref{eq:delta}).
In that limit, ${\cal R}$'s trajectory $(X_t,V_t)$ becomes identical for every realization of the process, therefore $\Delta\phi=0$, as in Eqs.~\ref{eq:Xevoln-na} - \ref{eq:dphi0}.
Additionally, the conditional dependence of $\Delta\sigma$ on ${\cal R}$'s initial and final states drops out of Eq.~\ref{eq:Deltasigma}.
Eq.~\ref{eq:sft-a} then becomes
\be
\label{eq:sft-na}
\left\langle e^{\beta Q - \Delta\sigma} \right\rangle = 1 \quad .
\ee
This result is the counterpart of Eqs.~4, 18 and 26 of Refs.~\cite{Crooks1999,Jarzynski1999a,Seifert2005}, which were derived for various models of an externally driven system in contact with one~\cite{Crooks1999,Seifert2005} or multiple~\cite{Jarzynski1999a} thermal baths.
Those results, in turn, followed earlier work on detailed fluctuation theorems for entropy production, for systems evolving in, or relaxing to, nonequilibrium steady states~\cite{Evans1993,Evans1994,Gallavotti1995a,Gallavotti1995b,Kurchan1998,Lebowitz1999a,Maes1999}.

\section*{Crooks's fluctuation theorem for autonomous work}

Crooks~\cite{Crooks1999} has derived a detailed fluctuation theorem for non-autonomous work:
\be
\label{eq:cft}
\frac{P^F(+W)}{P^R(-W)} = e^{\beta(W-\Delta F)} \quad .
\ee
Here, $P^F$ and $P^R$ are work distributions for {\it forward} and {\it reverse} non-autonomous processes.
These processes correspond to protocols, $\lambda_t^F$ and $\lambda_t^R$, for varying the work parameter from $t=0$ to $\tau$.
The protocols are related by time-reversal:
\be
\label{eq:FRsymmetry}
\lambda_t^F = \lambda_{\tau-t}^R \quad,\quad 0\le t \le\tau \quad .
\ee
In Eq.~\ref{eq:cft}, $\Delta F = F(\lambda_\tau^F) - F(\lambda_0^F)$.

For our purposes, it will be convenient to define the {\it dissipated work},
\be
W_{\rm diss} \equiv W - \left[ F(\lambda_\tau^i) - F(\lambda_0^i) \right] \quad,\quad i = F,R \quad .
\ee
In terms of this quantity, we rewrite Crooks's fluctuation theorem as
\be
\label{eq:cft-wdiss}
\frac{P^F(+W_{\rm diss})}{P^R(-W_{\rm diss})} = e^{\beta W_{\rm diss}} \quad .
\ee

In autonomous processes, the dynamical coordinate $X$ replaces the parameter $\lambda$.
It is not immediately obvious how to extend the notion of ``forward'' and ``reverse'' processes, defined by Eq.~\ref{eq:FRsymmetry}, to the autonomous setting, where $X$ evolves without external manipulation.
To obtain an autonomous version of Eq.~\ref{eq:cft}/\ref{eq:cft-wdiss}, we will consider two processes, labeled $F$ and $R$, characterized by identical dynamics but differently distributed initial conditions.
The connection to Crooks's forward and reverse processes will emerge only in the limit $M\rightarrow\infty$, combined with appropriate assumptions regarding initial distributions.

For either autonomous process, $F$ or $R$, we suppose that $\cal{RS}$ evolves from $t=0$ to $\tau$ under dynamics given by Eq.~\ref{eq:stochEOM}.
We now impose two assumptions that were not required in previous sections.
First, $H_{\cal S}$ is time-reversal-invariant:
\be
\label{eq:tri}
H_{\cal S}({\bf q},{\bf p};X) = H_{\cal S}({\bf q},-{\bf p};X) \quad .
\ee
Second, the propagator $K_t$ (Eq.~\ref{eq:propagator}) satisfies {\it detailed balance},
\be
\label{eq:db}
\frac{K_t(\Gamma^\prime \vert \Gamma)}{K_t(\Gamma^* \vert \Gamma^{\prime*})} =
\frac{ \exp [-\beta H_{\cal RS}(\Gamma^\prime)] }{ \exp [-\beta H_{\cal RS}(\Gamma)] } \quad ,
\ee
for all $t>0$.
Here and below, $(X,V,{\bf q},{\bf p})^* \equiv (X,-V,{\bf q},-{\bf p})$.
Equation~\ref{eq:db} is similar to Eq.~\ref{eq:wdbK}, but stronger.

Consider two probability distributions,
\begin{equation}
\label{eq:cft-a-ic}
f_0^F(\Gamma) = \rho_0^F(X,V) \, \pi({\bf z} \vert X)
\quad {\rm and} \quad 
f_0^R(\Gamma) = \rho_0^R(X,V) \, \pi({\bf z} \vert X) \quad ,
\end{equation}
where $\rho_0^{F,R}$ are distributions for ${\cal R}$, and Eq.~\ref{eq:pi} defines the conditional equilibrium distribution $\pi({\bf z}\vert X)$ for ${\cal S}$.
In processes $F$ and $R$, initial conditions for $\cal{RS}$ are sampled from $f_0^F$ and $f_0^R$, respectively.
Further define, for a trajectory $\Gamma_t$,
\begin{subequations}
\label{eq:Sigmadefs}
\begin{align}
\Sigma^F[\Gamma_t] &= \beta (W - \Delta F) -\ln \frac{ \rho_0^R(X_\tau,-V_\tau) }{ \rho_0^F(X_0,V_0) } \\
\Sigma^R[\Gamma_t] &= \beta (W - \Delta F) -\ln \frac{ \rho_0^F(X_\tau,-V_\tau) }{ \rho_0^R(X_0,V_0) } \quad ,
\end{align}
with $W[\Gamma_t]$ and $\Delta F[\Gamma_t]$ given by Eqs.~\ref{eq:Wstoch} and \ref{eq:DeltaF}.
\end{subequations}
To prevent the logarithmic terms from diverging, we assume -- for now -- that $\rho_0^F$ and $\rho_0^R$ are non-vanishing at all points $(X,V)$ in ${\cal R}$'s phase space; aside from this assumption, $\rho_0^F$ and $\rho_0^R$ are arbitrary.

Next, for $i \in \{F,R\}$, define
\be
P^i(\Sigma) \equiv \left\langle \delta (\Sigma-\Sigma^i) \right\rangle_i
\ee
where $\langle\cdots\rangle_i$ denotes an average over realizations of process $i$.
Thus $P^F(\Sigma)$ is the distribution of values of $\Sigma^F$ for process $F$, and $P^R(\Sigma)$ is defined analogously for process $R$.

Finally, define the {\it conjugate twin} $\Gamma_t^\dagger$ of a trajectory $\Gamma_t$ by the relation
\be
\label{eq:conjugateTwin}
\Gamma_t^\dagger = \Gamma_{\tau-t}^*
\quad,\quad 0\le t\le\tau \, .
\ee
Intuitively, the trajectory $\Gamma_t^\dagger$ is obtained from $\Gamma_t$ by ``running the movie backward''.
Equations~\ref{eq:DeltaF}, \ref{eq:Wstoch}, \ref{eq:Q} and \ref{eq:Sigmadefs} imply
\be
\label{eq:timeSymmetries-a}
\begin{split}
W[\Gamma_t^\dagger] = -W[\Gamma_t]
\quad&,\quad
\Delta F[\Gamma_t^\dagger] = -\Delta F[\Gamma_t]
\\
Q[\Gamma_t^\dagger] = -Q[\Gamma_t]
\quad&,\quad
\Sigma^R[\Gamma_t^\dagger] = -\Sigma^F[\Gamma_t] \, .
\end{split}
\ee
For a trajectory $\Gamma_t$ and its twin $\Gamma_t^\dagger$, Eq.~\ref{eq:db} gives
\be
\label{eq:Kratio}
\frac{K_\tau(\Gamma_\tau\vert\Gamma_0)}{K_\tau(\Gamma_\tau^\dagger\vert\Gamma_0^\dagger)} =
\frac{ \exp [-\beta H_{\cal RS}(\Gamma_\tau)] }{ \exp [-\beta H_{\cal RS}(\Gamma_0)] } =
e^{-\beta Q[\Gamma_t]} \, .
\ee

Since $W[\Gamma_t]$ and $\Delta F[\Gamma_t]$ are determined entirely by $\Gamma_0$ and $\Gamma_\tau$ (Eqs.~\ref{eq:Wstoch}, \ref{eq:DeltaF}), the same is true of $\Sigma^F[\Gamma_t]$ and $\Sigma^R[\Gamma_t]$.
We analyze $P^F(\Sigma)$ by taking an average over the joint distribution of $\Gamma_0$ and $\Gamma_\tau$:
\begin{align}
P^F(\Sigma) &= \int d\Gamma_0 \int d\Gamma_\tau \, f_0^F(\Gamma_0) \, K_\tau(\Gamma_\tau\vert\Gamma_0) \, \delta( \Sigma - \Sigma^F[\Gamma_t] ) \nonumber \\
&= \int d\Gamma_0 \int d\Gamma_\tau \, f_0^R(\Gamma_0^\dagger) \,
\frac{ \rho_0^F(X_0,V_0) }{ \rho_0^R(X_0^\dagger,V_0^\dagger) } \,
\frac{ \pi( {\bf z}_0 \vert X_0 ) }{ \pi( {\bf z}_0^\dagger \vert X_0^\dagger ) }
\nonumber\\
& \qquad\qquad \cdot K_\tau(\Gamma_\tau^\dagger\vert\Gamma_0^\dagger) e^{-\beta Q[\Gamma_t]} \, \delta( \Sigma - \Sigma^F[\Gamma_t] ) \nonumber \\
&= \int d\Gamma_\tau^\dagger \int d\Gamma_0^\dagger \, f_0^R(\Gamma_0^\dagger) \,
\frac{ \rho_0^F(X_\tau^\dagger,-V_\tau^\dagger) }{ \rho_0^R(X_0^\dagger,V_0^\dagger) } \,
\frac{ \pi( {\bf z}_\tau^\dagger \vert X_\tau^\dagger ) }{ \pi( {\bf z}_0^\dagger \vert X_0^\dagger ) }
\nonumber\\
& \qquad\qquad \cdot K_\tau(\Gamma_\tau^\dagger\vert\Gamma_0^\dagger) e^{\beta Q[\Gamma_t^\dagger]} \,
\, \delta( \Sigma + \Sigma^R[\Gamma_t^\dagger] ) \nonumber \\
&= \int d\Gamma_\tau^\dagger \int d\Gamma_0^\dagger \, f_0^R(\Gamma_0^\dagger) \,
K_\tau(\Gamma_\tau^\dagger\vert\Gamma_0^\dagger) \,
\frac{ \rho_0^F(X_\tau^\dagger,-V_\tau^\dagger) }{ \rho_0^R(X_0^\dagger,V_0^\dagger) } \,
\nonumber\\
& \qquad\qquad \cdot e^{-\beta (W[\Gamma_t^\dagger] - \Delta F[\Gamma_t^\dagger]} \,
\delta( \Sigma + \Sigma^R[\Gamma_t^\dagger] ) \nonumber \\
&= \int d\Gamma_\tau^\dagger \int d\Gamma_0^\dagger \, f_0^R(\Gamma_0^\dagger) \,
K_\tau(\Gamma_\tau^\dagger\vert\Gamma_0^\dagger) \,
\nonumber\\
& \qquad\qquad \cdot e^{-\beta\Sigma^R[\Gamma_t^\dagger]} \,
\delta( \Sigma + \Sigma^R[\Gamma_t^\dagger] ) \nonumber \\
&= e^{\Sigma} \, P^R(-\Sigma) \, ,
\end{align}
using Eqs.~\ref{eq:pi}, \ref{eq:Q}, \ref{eq:tri}, \ref{eq:cft-a-ic}, \ref{eq:Sigmadefs}, \ref{eq:conjugateTwin}-\ref{eq:Kratio}.
Equivalently,
\be
\label{eq:cft-a}
\frac{P^F(+\Sigma)}{P^R(-\Sigma)} = e^\Sigma \quad .
\ee

At this stage, $F$ and $R$ are mere labels.  Equation~\ref{eq:cft-a} is valid for arbitrary $\rho_0^{F,R}$, as long as both are non-vanishing everywhere.
To relate Eq.~\ref{eq:cft-a} to Crooks's fluctuation theorem, we first take the limit $M\rightarrow\infty$.
Then for every realization of either process, $F$ or $R$, the coordinate $X$ moves at constant velocity (Eq.~\ref{eq:Xevoln-na}).
Next, for a given choice of the distribution $\rho_0^F(X,V)$, we set
\be
\label{eq:FR-a}
\rho_0^R(X,V) = \rho_0^F(X+V\tau,-V) \, .
\ee
During process $F$, the work source evolves, at the ensemble level, from an initial distribution $\rho_0^F$ to a final distribution $\rho_\tau^F=\rho_0^{R*}$.
During process $R$ it evolves from $\rho_0^R$ to $\rho_\tau^R=\rho_0^{F*}$.
We see a connection to the intuitive notions of ``forward'' and ``reverse'' emerging.

At the trajectory level, Eq.~\ref{eq:FR-a} implies
\be
\label{eq:trajectoryLevel}
\rho_0^R(X_\tau,-V_\tau) = \rho_0^F(X_0,V_0)
\quad {\rm and} \quad
\rho_0^F(X_\tau,-V_\tau) = \rho_0^R(X_0,V_0)
\ee
for any trajectory $\Gamma_t$ generated by Eq.~\ref{eq:stochEOM} (with $M\rightarrow\infty$).
The logarithmic terms in Eq.~\ref{eq:Sigmadefs} now vanish, and Eq.~\ref{eq:timeSymmetries-a} gives
\be
\label{eq:wdiss-a}
\Sigma^F[\Gamma_t] = -\Sigma^R[\Gamma_t^\dagger] = \beta \left( W[\Gamma_t] - \Delta F[\Gamma_t] \right) \equiv W_{\rm diss}[\Gamma_t] \quad .
\ee
Equation~\ref{eq:cft-a} in turn becomes:
\be
\label{eq:cft-wdiss-a}
\frac{P^F(+W_{\rm diss})}{P^R(-W_{\rm diss})} = e^{\beta W_{\rm diss}} \quad .
\ee

Equations~\ref{eq:cft-wdiss} and \ref{eq:cft-wdiss-a} look the same, but they are not equivalent, since $\Delta F[\Gamma_t]$ in Eq.~\ref{eq:wdiss-a} fluctuates from one trajectory  to another.
To eliminate these fluctuations, let us assume delta-distributed initial conditions for process $F$.
Proceeding carefully, consider a family of well-behaved (everywhere non-vanishing) distributions $\rho_0^F(X,V;\epsilon)$, parametrized by $\epsilon>0$, such that
\be
\label{eq:epsilonLimit}
\lim_{\epsilon\rightarrow 0} \rho_0^F(X,V;\epsilon) = \delta(X-X_A) \, \delta(V-V_A) \, .
\ee
(For instance, $\rho_0^F$ could be a two-dimensional Gaussian that becomes infinitely narrow as $\epsilon\rightarrow 0$.)
Using Eq.~\ref{eq:FR-a}, define a corresponding family $\rho_0^R(X,V;\epsilon)$.
Equations~\ref{eq:FR-a} and \ref{eq:epsilonLimit} together imply
\be
\lim_{\epsilon\rightarrow 0} \rho_0^R(X,V;\epsilon) = \delta(X-X_B) \, \delta(V+V_A) \, ,
\ee
where $X_B = X_A + V_A \tau$.
For fixed $\epsilon>0$, when we take the limit $M\rightarrow\infty$ we obtain Eqs.~\ref{eq:trajectoryLevel} - \ref{eq:cft-wdiss-a}, as discussed above.
If we subsequently take the limit $\epsilon\rightarrow 0$, then we arrive at a situation in which the coordinate $X$ evolves from $X_A$ to $X_B$ for every realization of process $F$, and from $X_B$ to $X_A$ for every realization of process $R$.
Fluctuations in the value of $\Delta F$ are thereby eliminated, as they were in Eq.~\ref{eq:dF-noFluctuations}.
If we finally identify $X_A$ and $X_B$ in Eq.~\ref{eq:cft-wdiss-a} with $\lambda^F(0)$ and $\lambda^F(\tau)$ in Eq.~\ref{eq:cft-wdiss}, then the two equations become equivalent.

The preceding three paragraphs invoke two limits, taken in a particular order: first $M\rightarrow\infty$, then $\epsilon\rightarrow 0$.
If we had instead taken $\epsilon\rightarrow 0$ at fixed, finite $M$, then this limit would generically cause the logarithmic terms in Eqs.~\ref{eq:Sigmadefs} to diverge.
Therefore the ordering of limits is important, in the derivation presented above.

\section*{Discussion}

We have derived autonomous counterparts, Eqs.~\ref{eq:ft-a-intro}, \ref{eq:ft-exc}, \ref{eq:sft-a} and \ref{eq:cft-a}, of four key fluctuation theorems for non-autonomous work, Eqs.~\ref{eq:ft-na}, \ref{eq:ft-na-exc}, \ref{eq:sft-na} and \ref{eq:cft}.
We end with a brief discussion, including potential avenues of future research.

In the non-autonomous setting, we have invoked an external agent who manipulates a work parameter, $\lambda$, according to a predetermined schedule, or {\it protocol}.
It is useful to imagine the same agent in the autonomous setting.
Here the agent's role is to prepare the work source in its initial state, $\rho_0(X,V)$, and then to stand back and watch, without interfering, as ${\cal R}$ and ${\cal S}$ evolve in time.
From this perspective, the choice of ${\cal R}$'s initial state, $\rho_0(X,V)$, in the autonomous setting, is the counterpart of the choice of protocol, $\lambda_t$, in the non-autonomous setting.
While the time-dependence of $\lambda$ is fully determined in the latter case, the evolution of $X$ is only partially determined (through the choice of initial conditions) in the former case.

Our central results are readily generalized to include a potential energy term, $U(X)$, that acts on the work source.
In this case, we have $H_{\cal R}^0=MV^2/2 + U(X)$ in place of $H_{\cal R}^0=MV^2/2$ (see Eq.~\ref{eq:HRS}); $W =  H_{\cal R}^0(X_0,V_0) - H_{\cal R}^0(X_\tau,V_\tau)$ in Eqs.~\ref{eq:W} and \ref{eq:Wstoch}; and in the exclusive convention $H_{\cal R}$ includes the potential term $U$.
With these modifications, the derivations appearing in Eqs.~\ref{eq:ft}, \ref{eq:ft-a-rst}, \ref{eq:ft-exc}, \ref{eq:sft-a} and \ref{eq:cft-a} remain essentially unchanged, for any finite value of $M$.
The infinite-inertia limit requires some care.
If $U(X)$ does not explicitly depend on $M$, then Eq.~\ref{eq:Xevoln-na} remains valid as $M\rightarrow\infty$: the massive work source moves at constant velocity.
If, however, the potential is proportional to ${\cal R}$'s mass, e.g.\ $U(X)=M\Phi(X)$, then in the limit $M\rightarrow\infty$, ${\cal R}$'s trajectory obeys $\ddot X_t = -\partial_X\Phi(X_t)$, without back-action.
(For instance, if ${\cal R}$ is a mass in a uniform gravitational field, then $U=MgX$ and $X_t=X_0+V_0t+gt^2/2$ in the infinite-inertia limit.)
In this case Eqs.~\ref{eq:rhoShear} and \ref{eq:FR-a} are suitably modified to reflect a nonlinear trajectory $X_t$, but otherwise the arguments we have made about the infinite-inertia limit are unchanged.

Do our results extend to the quantum realm?
To address this question, we must first tackle a more basic one:
how are work and heat to be defined for quantum systems undergoing thermodynamic processes?
For closed quantum systems, work is most commonly defined via the {\it two-point measurement} scheme, which consists of projective measurements of ${\cal S}$'s initial and final energies~\cite{Tasaki2000,Kurchan2000,Mukamel2003,Esposito2009,Campisi2011} -- the intuition is that in the absence of a thermal reservoir, the work performed on ${\cal S}$ is the change in its internal energy. This notion of work, however, suffers from quantum back-action \cite{Deffner2016PRE}, which would have to be accounted for in addition to the classical back-action discussed above.
Viable alternatives to the two-point scheme have been suggested, for closed quantum systems \cite{Deffner2016PRE,Landi2020PRL,Maeda2023PRA}.
For open quantum systems, the situation is even more involved and how to define work and heat remains an unresolved question and a subject of active research~\cite{Campisi2009,Deffner2011EPL,Deffner2011PRL,Kafri2012PRA,Rastegin2013,Deffner2013EPL,Rastegin2014,SantosPRL2017,Miller2019,Strasberg2019,Rivas2020,Sone2023AVS,Jacob2024,Davoudi2025}.

The idea of defining work autonomously -- in terms of dynamically evolving work reservoirs -- appears more frequently in the quantum thermodynamics literature than in classical stochastic thermodynamics.
Refs.~\cite{Cohen2012,Alhambra2016,Monsel2018,Lipka-Bartosik2021,Lyu20204,Han2024,Han2025,Varma2025} derive and explore a variety of quantum fluctuation theorems, for which work is taken to be the energy transferred from a work reservoir to a system of interest.
In particular, Cohen and colleagues~\cite{Han2024,Varma2025} take an approach that emphasizes the effects of back-action on the quantum work reservoir.
It will be interesting to clarify how the quantum fluctuation theorems of Refs.~\cite{Cohen2012,Alhambra2016,Monsel2018,Lipka-Bartosik2021,Lyu20204,Han2024,Han2025,Varma2025} relate to the classical ones we have obtained.

Our results may be relevant for the thermodynamics of feedback control.
A generic measurement-and-feedback setting features a {\it controller}, ${\cal C}$.
The controller performs measurements on the evolving system of interest, ${\cal S}$, and manipulates it based on  measurement outcomes.
If the controller is a physical device -- a gadget or robot -- then feedback control emerges dynamically from the interplay between system and controller:
${\cal S}$'s evolution influences ${\cal C}$'s evolution, which in turn influences ${\cal S}$ in a way that depends on how ${\cal S}$ was evolving in the first place.
From this perspective, measurement and feedback are no more than an intricate form of back-action, and the controller ${\cal C}$ is analogous to our work source, ${\cal R}$.
It would be interesting to investigate whether fluctuation theorems for feedback control, such as those obtained by Sagawa and Ueda~\cite{Sagawa2010}, can be understood within the autonomous framework that we have developed.
Hartich, Barato and Seifert~\cite{Hartich2014}, and Horowitz and Esposito~\cite{Horowitz2014} have explored closely related ideas for discrete-state, bipartite systems.
Ehrich and Sivak~\cite{Ehrich2023} have emphasized the relevance of bipartite Markovian dynamics to biomolecular sensors.

More generally, when two physical systems interact with one another, each one influences and is influenced by the other.  Is there a natural way to tease apart which one is performing measurement and feedback, and which one is being measured and manipulated?
Or perhaps a useful way to quantify the degree to which each system is performing measurement and feedback on the other?

To illustrate the general results we have derived, it would be useful and satisfying to identify models of ${\cal R}$ and ${\cal S}$ whose dynamics, whether Hamiltonian or stochastic, can be solved exactly.

Ultimately, it would be desirable to realize our predictions experimentally.
A potential obstacle is the requirement that the work source, ${\cal R}$, must not be in direct contact with the thermal reservoir, ${\cal T}$.
This obstacle vanishes if there is no reservoir at all, that is if ${\cal RS}$ evolves under Hamiltonian dynamics, Eq.~\ref{eq:hameq}.
In this scenario, one can imagine an experiment using a macroscopic system with (at least) two degrees of freedom, such as a pendulum swinging in two dimensions.
One degree would be the system of interest, ${\cal S}$, the other the work source, ${\cal R}$.
Initial conditions sampled from the statistical state $\rho_0 \pi$ (Eq.~\ref{eq:f0}) could be generated randomly by computer, at a fictitious temperature $\beta^{-1}$.\footnote{
In order for the (macroscopic) system's fluctuations to be non-negligible, this temperature would need to be absurdly large.
}
The pendulum would need to be sufficiently dissipation-free for its dynamics to be approximately Hamiltonian over the duration $\tau$ of the process.

Now imagine instead that ${\cal S}$ is a microscopic system immersed in a thermal reservoir.
For specificity, consider a micron-size polystyrene bead in water, trapped harmonically by optical tweezers.
It would be difficult to couple this bead to another microscopic system, playing the role of the work source, ${\cal R}$, without having the latter also immersed in the reservoir.
Suppose, however, that we treat the position $X$ of the optical trap (along some direction) as a work source, so that the system Hamiltonian $H_{\cal S}({\bf z};X)$ describes a harmonic oscillator.
Suppose further that we assign an arbitrary mass $M$ to the work source.
Then the dynamics given by Eq.~\ref{eq:stochEOM} could be achieved experimentally by (1) allowing ${\cal S}$ to evolve under the stochastic dynamics generated by the bead's interactions with the trap and the surrounding water, while (2) using measurement and feedback to move the trap's position $X$ along a trajectory that satisfies $M\ddot X = - \partial H_{\cal S}/\partial X$.
In this scenario, the autonomous evolution of ${\cal RS}$ is mocked up by externally imposing Newton's equations of motion on the trap's position.
In effect, the back-action of ${\cal S}$ on ${\cal R}$ is put in by hand.
Such an experiment would require fast and accurate measurement and feedback capabilities.
An advantage of this approach is that the work source's mass, $M$, is a tunable parameter, allowing the limit $M \rightarrow\infty$ to be probed experimentally.

\acknow{C.J. acknowledges support from the U.S. National Science Foundation under Grant No. DMR-2127900, and from the Simons Foundation under Award No. 681566. S.R. acknowledges support from the Israel Science Foundation under Grant No. 1929/21. S.D. acknowledges support from the John Templeton Foundation under Grant No. 62422.
}

\showacknow{} 

\bibsplit[42]

\bibliography{refs}

\newpage
\appendix
\onecolumn






\renewcommand{\thesection}{\arabic{section}}
\setcounter{section}{0}

\renewcommand{\theequation}{S\arabic{equation}}
\setcounter{equation}{0}



\centerline{
{\Large {\bf Supporting Information for ``Fluctuation theorems for autonomous work''}} }
\vskip .1in

\centerline{
{\large {\bf Christopher Jarzynski, Sebastian Deffner and Saar Rahav}} }
\vskip .2in

In the main text, when deriving autonomous fluctuation theorems in the presence of a heat bath, the bath was modeled implicitly, using stochastic equations of motion (Eq.~21).
Here we derive the same fluctuation theorems -- Eqs.~30, 44, 50 and 64 -- within a Hamiltonian framework that treats the bath's microscopic degrees of freedom explicitly.
As in the main text, we consider a work source, $\cal R$, and a system of interest, $\cal S$, whose microscopic states are denoted by $(X,V)$ and $\bf z$, respectively.
We additionally consider a heat bath, $\cal T$, whose microstate is denoted by ${\bf y}$.
Our ``universe'' thus consists of the combined system ${\cal RST}$, whose microstate is  $\Gamma=(X,V,{\bf y},{\bf z})$.

The total energy of ${\cal RST}$ is given by the Hamiltonian
\be
\label{eq:HRST}
H_{\cal RST}(\Gamma) = \frac{M}{2} V^2 + H_{\cal S}({\bf z};X) + H_{\cal T}({\bf y}) + h_{\rm int}({\bf y},{\bf z}) \, .
\ee
The terms $MV^2/2$ and $H_{\cal S}$ have the same meaning as in the main text;
$H_{\cal T}$ is the heat bath Hamiltonian; and $h_{\rm int}$ is an interaction term that couples ${\cal S}$ and ${\cal T}$. 
By assumption, ${\cal S}$ interacts with both ${\cal R}$ and ${\cal T}$ (via $H_{\cal S}$ and $h_{\rm int}$), but ${\cal R}$ and ${\cal T}$ do not interact directly with one another.
This assumption reflects standard idealizations about a work source and a heat bath.

Throughout this appendix, we assume that ${\cal RST}$ evolves under Hamilton's equations, generated by $H_{\cal RST}(\Gamma)$.
This evolution is specified by a deterministic trajectory $\Gamma_t$, with $0\le t\le\tau$.
The value of $H_{\cal RST}$ is conserved along this trajectory.

Defining
\ba
\label{eq:HR0}
H_{\cal R}^0(V) &=& \frac{M}{2} V^2 \\
\label{eq:HST}
H_{\cal ST}({\bf y},{\bf z};X) &=& H_{\cal S}({\bf z};X) + H_{\cal T}({\bf y}) + h_{\rm int}({\bf y},{\bf z}) \, ,
\ea
we have
\be
H_{\cal RST} = H_{\cal R}^0 + H_{\cal ST} \, .
\ee
Throughout most of this Supporting Information (Secs.~\ref{sec:WQ}, \ref{sec:ft-inc}, \ref{sec:entropy}, \ref{sec:cft}), we consider $H_{\cal R}^0$ to be the energy of the work source, and $H_{\cal ST}$ to be that of the combined system of interest and heat bath.
As in the main text, this convention for assigning energies to subsystems corresponds to the inclusive definition of work~\cite{Jarzynski2007}.
In Sec.~\ref{sec:ft-exc}, where we derive a fluctuation theorem for exclusive work, we use a different convention (Eq.~\ref{eq:exclusiveSplit}).

In the main text, the conditional equilibrium state of ${\cal S}$, Eq.~9, played an important role.
Here we will need the conditional equilibrium state of ${\cal ST}$:
\be
\label{eq:Pi}
\begin{split}
\Pi({\bf y},{\bf z} \vert X) &= \frac{1}{Z_{\cal ST}(X)} e^{-\beta H_{\cal ST}({\bf y},{\bf z}; X)} \\
Z_{\cal ST}(X) &= \int d{\bf y} \int d{\bf z} \, e^{-\beta H_{\cal ST}({\bf y},{\bf z}; X)} \, .
\end{split}
\ee
The probability density $\Pi({\bf y},{\bf z} \vert X)$ represents the equilibrium state of ${\cal ST}$, conditioned on a fixed value of ${\cal R}$'s coordinate, $X$.
To obtain the conditional equilibrium state of ${\cal S}$ alone, we integrate over the microstate of ${\cal T}$:
\be
\label{eq:pi-SI}
\pi({\bf z} \vert X) = \int d{\bf y} \, \Pi({\bf y},{\bf z}\vert X) = \frac{1}{Z_{\cal S}^+(X)} e^{-\beta H_{\cal S}^+({\bf z};X)} \, ,
\ee
where
\be
\begin{split}
\label{eq:HMF}
H_{\cal S}^+({\bf z};X) &= H_S({\bf z};X) + \Phi({\bf z}) \\
\Phi({\bf z}) &= -\beta^{-1} \ln \frac{ \int d{\bf y} \, e^{ -\beta \left[ H_{\cal T}({\bf y}) + h_{\rm int}({\bf y},{\bf z}) \right] } }{ \int d{\bf y} \, e^{ -\beta H_{\cal T}({\bf y}) } } \\
Z_{\cal S}^+(X) &= \int d{\bf z} \, e^{-\beta H_{\cal S}^+({\bf z}; X) } \, .
\end{split}
\ee
$H_{\cal S}^+({\bf z};X)$ is the system of interest's {\it Hamiltonian of mean force}~\cite{Kirkwood1935}, conditioned on $X$.
This quantity plays a central role in the classical statistical physics of systems that are coupled strongly to their environments~\cite{Roux1999,Jarzynski2004,Gelin2009,Seifert2016,Talkner2016,Jarzynski2017}.
We see in Eq.~\ref{eq:pi-SI} that $H_{\cal S}^+ \, (= H_{\cal S} + \Phi)$ acts as an effective Hamiltonian.
The term $\Phi({\bf z})$ describes how the system-bath interaction, $h_{\rm int}$, affects the equilibrium state of ${\cal S}$.
When $h_{\rm int}$ is sufficiently small, this ``correction term'' becomes negligible ($\Phi\rightarrow 0$), and Eq.~\ref{eq:pi-SI} reduces to the familiar Boltzmnn-Gibbs expression, as in Eq.~9 of the main text.
In this appendix we do not assume that either $h_{\rm int}$ or $\Phi$ is negligible.

Defining
\be
\label{eq:zt0}
Z_{\cal T}^0 = \int d{\bf y} \, e^{-\beta H_{\cal T}({\bf y})} \, ,
\ee
Eqs.~\ref{eq:Pi} and \ref{eq:pi-SI} give us the useful identity
\be
\label{eq:Zidentity}
Z_{\cal ST}(X) = Z_{\cal S}^+(X) \, Z_{\cal T}^0 \, .
\ee
Following Refs.~\cite{Roux1999,Jarzynski2004,Gelin2009,Seifert2016,Talkner2016,Jarzynski2017}, we take the free energy of ${\cal S}$, conditioned on ${\cal R}$'s coordinate, to be
\be
\label{eq:F}
F(X)  = -\beta^{-1} \ln Z_{\cal S}^+(X) \, .
\ee

\section{Work and heat}
\label{sec:WQ}


As ${\cal RST}$ evolves along a trajectory $\Gamma_t$, the values of $H_{\cal R}^0$ and $H_{\cal ST}$ change with time but their sum, $H_{\cal RST}$, remains fixed.
The energy lost by ${\cal R}$ is the work performed on ${\cal S}$:
\be
\label{eq:energyConserv}
W = -\Delta H_{\cal R}^0 = \Delta H_{\cal ST} \, .
\ee
The notation $\Delta$ indicates the net change in the value of a quantity, over the time interval from $t=0$ to $\tau$.
In more detail,
\ba
W[\Gamma_t] = \frac{M}{2}\left( V_0^2 - V_\tau^2 \right) &=& H_{\cal ST}({\bf y}_\tau,{\bf z}_\tau;X_\tau) - H_{\cal ST}({\bf y}_0,{\bf z}_0;X_0) \nonumber \\
&=& \int_0^\tau dt \, \dot \Gamma_t \frac{ \partial H_{\cal ST} }{ \partial \Gamma }(\Gamma_t) \nonumber \\
&=& \int_0^\tau dt \, \dot X_t \frac{ \partial H_{\cal ST} }{ \partial X }({\bf z}_t;X_t) \nonumber \\
\label{eq:work}
&=& \int_0^\tau dt \, \dot X_t \frac{ \partial H_{\cal S} }{ \partial X }({\bf z}_t;X_t) = \int_0^\tau dt \, \dot X_t \frac{ \partial H_{\cal S}^+ }{ \partial X }({\bf z}_t;X_t) \, .
\ea
We use Hamilton's equations to get to the third line, and Eqs.~\ref{eq:HST} and \ref{eq:HMF} to get to the fourth line.

Since $H_{\cal S}^+$ is an effective Hamiltonian in the Boltzmann-like factor appearing in Eq.~\ref{eq:pi-SI}, we interpret $H_{\cal S}^+$ as the internal energy of ${\cal S}$~\cite{Jarzynski2004}.
By the first law of thermodynamics, the net change in this internal energy is a sum of two contributions, work and heat:
\be
\label{eq:firstlaw}
\Delta H_{\cal S}^+ = W + Q \, .
\ee
Eqs.~\ref{eq:firstlaw} and \ref{eq:work} lead to the following expression for the heat absorbed by ${\cal S}$:~\footnote{
Eq.~\ref{eq:heat} is equivalent to Eq.~(35a) of Ref.~\cite{Jarzynski2017}, only there $H_{\cal S}^+$ (which appears as $h_{\cal S}$ in Ref.~\cite{Jarzynski2017}) is interpreted as {\it enthalpy} rather than internal energy.
The distinction is not relevant for the present paper.
}
\be
\label{eq:heat}
Q[\Gamma_t] = \int_0^\tau dt \, \dot {\bf z}_t \frac{ \partial H_{\cal S}^+ }{ \partial {\bf z} }({\bf z}_t;X_t) \, .
\ee
Eqs.~\ref{eq:work} and \ref{eq:heat} have the same structure as the standard expressions for work and heat in stochastic thermodynamics~\cite{Seifert2012,Peliti2021}, except that the Hamiltonian of mean force takes the place of the system Hamiltonian, and the coordinate $X$ takes the place of the work parameter.

The interpretation of $H_{\cal S}^+$ as internal energy, and the corresponding definition of heat, Eq.~\ref{eq:heat}, are used in Sec.~\ref{sec:entropy} when deriving a fluctuation theorem for total entropy production, but nowhere else in this Supporting Information.

We are now in a position to derive the central results of the main text, Eqs.~30, 44, 50 and 64, within a framework that explicitly treats the degrees of freedom of the work source, system of interest and heat bath.

\section{Fluctuation theorem for inclusive work}
\label{sec:ft-inc}

We first suppose that initial conditions for ${\cal RST}$ are sampled from
\be
\label{eq:f0-SI}
f_0(\Gamma) = \rho_0(X,V) \, \Pi({\bf y},{\bf z} \vert X) \, ,
\ee
where $\rho_0(X,V)$ is an arbitrary distribution on ${\cal R}$'s phase space, and ${\cal ST}$ is in a state of conditional equilibrium (Eq.~\ref{eq:Pi}).
Eq.~\ref{eq:f0-SI} is analogous to Eq.~10 of the main text.
From these initial conditions, ${\cal SRT}$ evolves along a Hamiltonian trajectory $\Gamma_t$.
Let $f_t(\Gamma)$ be the time-evolving distribution that describes an ensemble of these trajectories, and let $\rho_t(X,V) = \int d{\bf y} \int d{\bf z} \, f_t(\Gamma)$ denote the marginal distribution of ${\cal R}$.
We define the free energy change of ${\cal S}$, and the stochastic entropy change of ${\cal R}$, as in the main text:
\begin{align}
\label{eq:DeltaF-SI}
\Delta F[\Gamma_t] &= F(X_\tau) - F(X_0)  \\ 
\label{eq:Deltaphi-SI}
\Delta\phi[\Gamma_t] &= -\ln\rho_\tau(X_\tau,V_\tau) + \ln\rho_0(X_0,V_0) \, ,
\end{align}
with $F(X)$ given by Eq.~\ref{eq:F}.

Since $\Gamma_t$ evolves deterministically, $W$, $\Delta F$ and $\Delta\phi$ are determined uniquely by initial conditions $\Gamma_0$.
We then have
\ba
\left\langle e^{-\beta(W-\Delta F)-\Delta\phi} \right\rangle
&=& \int d\Gamma_0 \, f_0(\Gamma_0) \, e^{-\beta(W-\Delta F)-\Delta\phi} \nonumber \\
&=& \int d\Gamma_0 \, f_0(\Gamma_0) \, \frac{e^{-\beta H_{\cal ST}({\bf y}_\tau,{\bf z}_\tau;X_\tau)} }{ e^{-\beta  H_{\cal ST}({\bf y}_0,{\bf z}_0;X_0)} }
\frac{Z_{\cal S}^+(X_0)}{Z_{\cal S}^+(X_\tau)}
\frac{ \rho_\tau(X_\tau,V_\tau) }{ \rho_0(X_0,V_0) } \nonumber\\
&=& \int d\Gamma_0 \, f_0(\Gamma_0) \, \frac{ Z_{\cal ST}(X_\tau) \Pi({\bf y}_\tau,{\bf z}_\tau \vert X_\tau) }{ Z_{\cal ST}(X_0) \Pi({\bf y}_0,{\bf z}_0 \vert X_0) }
\frac{Z_{\cal S}^+(X_0)}{Z_{\cal S}^+(X_\tau)}
\frac{ \rho_\tau(X_\tau,V_\tau) }{ \rho_0(X_0,V_0) } \nonumber \\
\label{eq:ft-a-rst-explicit}
&=& \int d\Gamma_\tau \, \rho_\tau(X_\tau,V_\tau) \, \Pi({\bf y}_\tau,{\bf z}_\tau \vert X_\tau) = 1 \, .
\ea
We used Eqs.~\ref{eq:F}, \ref{eq:energyConserv}, \ref{eq:DeltaF-SI} and \ref{eq:Deltaphi-SI} to get to the second line, Eq.~\ref{eq:Pi} to get to the third line, and Eqs.~\ref{eq:Zidentity} and \ref{eq:f0-SI} to get to the fourth line.

Eq.~\ref{eq:ft-a-rst-explicit} is the counterpart of Eq.~30 of the main text, only here the result is obtained by modeling the microscopic degrees of ${\cal T}$ explicitly rather than implicitly.

\section{Fluctuation theorem for exclusive work}
\label{sec:ft-exc}

Following Eq.~38 of the main text, we partition $H_{\cal S}({\bf z};X)$ into a term that depends only on the microstate of ${\cal S}$, and an ${\cal R}$-${\cal S}$ interaction term:
\be
H_{\cal S}({\bf z};X) = H_{\cal S}^0({\bf z}) + H_{\rm int}({\bf z},X) \, .
\ee
We then write
\be
\label{eq:exclusiveSplit}
\begin{split}
H_{\cal RST}(\Gamma) &= \left[ \frac{M}{2} V^2 + H_{\rm int}({\bf z},X) \right] + \left[ H_{\cal S}^0({\bf z}) + H_{\cal T}({\bf y}) + h_{\rm int}({\bf y},{\bf z}) \right] \\
&= H_{\cal R}(X,V ; {\bf z}) + H_{\cal ST}^0({\bf y},{\bf z}) \, ,
\end{split}
\ee
and we treat $H_{\cal R}$ and $H_{\cal ST}^0$ as the energies of ${\cal R}$ and ${\cal ST}$, respectively.
This reassignment of subsystem energies corresponds to the {\it exclusive} convention, in contrast with the inclusive convention adopted in the previous and following sections.
In the exclusive convention, the ${\cal R}$-${\cal S}$ interaction energy contributes to the internal energy of ${\cal R}$ rather than ${\cal S}$, and the work performed on ${\cal S}$ is:
\be
\label{eq:W0-SI}
W^0 = -\Delta H_{\cal R} = \Delta H_{\cal ST}^0 \, .
\ee
The equilibrium state of ${\cal ST}$, no longer conditioned on $X$, is given by the distribution
\be
\label{eq:Pi0}
\Pi^0({\bf y},{\bf z}) = \frac{1}{Z_{\cal ST}^0} e^{-\beta H_{\cal ST}^0({\bf y},{\bf z})} \, .
\ee

We now suppose that initial conditions are sampled from 
\be
\label{eq:f00-SI}
f_0(\Gamma) = \rho_0(X,V) \, \Pi^0({\bf y},{\bf z}) \, ,
\ee
and we compute
\ba
\left\langle e^{-\beta W^0 - \Delta\phi} \right\rangle
&=& \int d\Gamma_0 \, f_0(\Gamma_0) \, e^{-\beta W^0 - \Delta\phi} \nonumber \\
&=& \int d\Gamma_0 \, f_0(\Gamma_0) \, 
\frac{e^{-\beta H_{\cal ST}^0({\bf y}_\tau,{\bf z}_\tau)} }{ e^{-\beta  H_{\cal ST}^0({\bf y}_0,{\bf z}_0)} }
\frac{ \rho_\tau(X_\tau,V_\tau) }{ \rho_0(X_0,V_0) } \nonumber \\
&=& \int d\Gamma_0 \, \rho_0(X_0,V_0) \, \Pi^0({\bf y}_0,{\bf z}_0)
\frac{\Pi^0({\bf y}_\tau,{\bf z}_\tau) }{ \Pi^0({\bf y}_0,{\bf z}_0) }
\frac{ \rho_\tau(X_\tau,V_\tau) }{ \rho_0(X_0,V_0) } \nonumber \\
\label{eq:ft-exc-explicit}
&=& \int d\Gamma_\tau \, \rho_\tau(X_\tau,V_\tau) \, \Pi^0({\bf y}_\tau,{\bf z}_\tau) = 1 \, ,
\ea
using Eqs.~\ref{eq:Deltaphi-SI}, \ref{eq:W0-SI}, \ref{eq:Pi0} and \ref{eq:f00-SI}.

Eq.~\ref{eq:ft-exc-explicit} is the counterpart of Eq.~44 of the main text.

\section{Fluctuation theorem for total entropy production}
\label{sec:entropy}

Using Eq.~\ref{eq:zt0}, let us rewrite Eq.~\ref{eq:HMF} as follows:
\be
\label{eq:HMF-rewrite}
H_{\cal S}^+({\bf z};X) = H_{\cal S}({\bf z};X) - \beta^{-1} \ln \frac{ Z_{\cal T}^+({\bf z}) }{ Z_{\cal T}^0 } \, ,
\ee
with
\be
Z_{\cal T}^+({\bf z}) = \int d{\bf y} \, e^{-\beta[H_{\cal T}({\bf y}) + h_{\rm int}({\bf y},{\bf z})] } \, .
\ee
We further introduce
\be
\label{eq:pit}
\pi_{\cal T}({\bf y}\vert{\bf z}) = \frac{1}{Z_{\cal T}^+({\bf z})} \, e^{-\beta[H_{\cal T}({\bf y}) + h_{\rm int}({\bf y},{\bf z})] } \, ,
\ee
which is the equilibrium state of ${\cal T}$, conditioned on the microstate of ${\cal S}$.

In Sections \ref{sec:ft-inc} and \ref{sec:ft-exc}, ${\cal R}$'s initial state was sampled from an arbitrary distribution $\rho_0$, while ${\cal ST}$'s was sampled from equilibrium (Eqs.~\ref{eq:f0-SI}, \ref{eq:f00-SI}).
Now we relax this assumption and allow the initial conditions of both ${\cal R}$ and ${\cal S}$ to be sampled arbitrarily, though we still take ${\cal T}$ to be sampled from equilibrium.
Specifically, initial conditions for ${\cal RST}$ are sampled from 
\be
\label{eq:f0-seifert}
f_0(\Gamma) = \rho_0(X,V) \, \eta_0({\bf z} \vert X,V) \, \pi_{\cal T}({\bf y}\vert{\bf z}) \, ,
\ee
where $\rho_0\eta_0$ represents a general distribution on $(X,V,{\bf z})$-space.
Letting $f_t(\Gamma)$ denote the distribution of ${\cal RST}$ a later time $t$, we define
\be
\begin{split}
\rho_t(X,V) &= \int d{\bf z} \int d{\bf y} \, f_t(X,V,{\bf y},{\bf z}) \\
\eta_t({\bf z} \vert X,V) &= \frac{ \int d{\bf y} \, f_t(X,V,{\bf y},{\bf z}) }{ \rho_t(X,V) } \, .
\end{split}
\ee
These expressions represent the marginal distributions of ${\cal R}$, and of ${\cal S}$ conditioned on ${\cal R}$, at time $t$.
The counterparts of these distributions were defined by Eq.~47 of the main text, where ${\cal T}$ was treated implicitly.

Along a trajectory $\Gamma_t$, the net changes in $H_{\cal S}^+$ and $H_{\cal S}$ are related by
\be
\label{eq:dH+dH}
\Delta H_{\cal S}^+ = \Delta H_{\cal S} - \beta^{-1} \ln \frac{ Z_{\cal T}^+({\bf z}_\tau) }{ Z_{\cal T}^+({\bf z}_0) } \, ,
\ee
using Eq.~\ref{eq:HMF-rewrite}.
We then use Eqs.~\ref{eq:firstlaw}, \ref{eq:energyConserv}, \ref{eq:pit} and \ref{eq:dH+dH}, along with the conservation of total energy, $\Delta H_{\cal RST}=0$, to obtain the identity
\ba
\frac{ \pi_{\cal T}({\bf y}_\tau\vert{\bf z}_\tau) }{ \pi_{\cal T}({\bf y}_0\vert{\bf z}_0) } &=& \frac{ Z_{\cal T}^+({\bf z}_0) }{ Z_{\cal T}^+({\bf z}_\tau) } \,
e^{-\beta ( \Delta H_{\cal T} + \Delta h_{\rm int} )} \nonumber \\
&=&
e^{\beta(\Delta H_{\cal S}^+ - \Delta H_{\cal S})} \, e^{\beta(\Delta H_{\cal R}^0 + \Delta H_{\cal S})} \nonumber \\
&=&
\label{eq:identity}
e^{\beta\Delta H_{\cal S}^+} \, e^{\beta\Delta H_{\cal R}^0} = e^{\beta(W+Q)} \, e^{-\beta W} = e^{\beta Q} \, .
\ea
Finally, defining $\Delta\phi = -\Delta\ln\rho$ and $\Delta\sigma = -\Delta\ln\eta$ as in the main text (Eqs.~14 and 48), and sampling initial conditions from $f_0$ (Eq.~\ref{eq:f0-seifert}), we use Eq.~\ref{eq:identity} to compute the average
\ba
\left\langle e^{ \beta  Q - \Delta\phi - \Delta\sigma } \right\rangle
&=& \int d\Gamma_0 \, f_0(\Gamma_0) \, \frac{ \pi_{\cal T}({\bf y}_\tau\vert{\bf z}_\tau) }{ \pi_{\cal T}({\bf y}_0\vert{\bf z}_0) } \,
\frac{ \rho_\tau(X_\tau,V_\tau) }{ \rho_0(X_0,V_0) }
\frac{ \eta_\tau({\bf z}_\tau \vert X_\tau,V_\tau) }{ \eta_0({\bf z}_0 \vert X_0,V_0) } \nonumber \\
\label{eq:sft-a-explicit}
&=&
\int d\Gamma_\tau \, \pi_{\cal T}({\bf y}_\tau\vert{\bf z}_\tau) \, \rho_\tau(X_\tau,V_\tau) \, \eta_\tau({\bf z}_\tau \vert X_\tau,V_\tau) = 1 \, .
\ea

Eq.~\ref{eq:sft-a-explicit} is the counterpart of Eq.~50 of the main text.

\section{Crooks's fluctuation theorem for autonomous work}
\label{sec:cft}

As in the main text, we consider two autonomous processes, $F$ and $R$, corresponding to two choices for initial conditions, $f_0^F(\Gamma)$ and $f_0^R(\Gamma)$;
see Eq.~58.
The microstate $\Gamma$ now includes the coordinates and momenta of the heat bath, ${\cal T}$.
We assume that ${\cal ST}$ begins in a conditional equilibrium state:
\be
\label{eq:f0FR}
\begin{split}
f_0^F(\Gamma) &= \rho_0^F(X,V) \,\Pi({\bf y},{\bf z} \vert X) \\
f_0^R(\Gamma) &= \rho_0^R(X,V) \,\Pi({\bf y},{\bf z} \vert X) \, .
\end{split}
\ee
From initial conditions $\Gamma_0$, sampled from either $f_0^F$ or $f_0^R$, a trajectory $[\Gamma_t]$ evolves under the Hamiltonian $H_{\cal RST}$.
For the two processes, we define $\Sigma^F$ and $\Sigma^R$ as in the main text (Eq.~59):
\be
\label{eq:SigmaFR}
\begin{split}
\Sigma^F[\Gamma_t] &= \beta(W-\Delta F) - \ln \frac{ \rho_0^R(X_\tau,-V_\tau) }{ \rho_0^F(X_0,V_0) } \\
\Sigma^R[\Gamma_t] &= \beta(W-\Delta F) - \ln \frac{ \rho_0^F(X_\tau,-V_\tau) }{ \rho_0^R(X_0,V_0) }
\end{split}
\ee
Using Eqs.~\ref{eq:Pi}, \ref{eq:F} and the first line of Eq.~\ref{eq:work}, we rewrite $\Sigma^R$ as
\be
\label{eq:SigmaR-convenient}
\Sigma^R[\Gamma_t] = -\ln \left[ \frac{ \rho_0^F(X_\tau,-V_\tau) }{ \rho_0^R(X_0,V_0) } \frac{ \Pi({\bf y}_\tau,{\bf z}_\tau \vert X_\tau) }{ \Pi({\bf y}_0,{\bf z}_0 \vert X_0) } \right] \, ,
\ee
which will prove convenient.

Analogously to Eq.~56, we assume $H_{\cal ST}$ is time-reversal-invariant,
\be
\label{eq:tri-SI}
H_{\cal ST}({\bf y}^*,{\bf z}^*;X) = H_{\cal ST}({\bf y},{\bf z};X) \, ,
\ee
which in turn implies
\be
\label{eq:Pi-tri}
\Pi({\bf y}^*,{\bf z}^* \vert X) = \Pi({\bf y},{\bf z} \vert X) \, .
\ee
From Eqs.~\ref{eq:f0FR}-\ref{eq:Pi-tri} we obtain the symmetry relation
\be
\label{eq:Sigma-sym}
\Sigma^F[\Gamma_t] = -\Sigma^R[\Gamma_t^\dagger] \, ,
\ee
where the trajectory $\Gamma_t^\dagger$ is the conjugate twin of the trajectory $\Gamma_t$:
\be
\label{eq:conjTwin}
\Gamma_t^\dagger = \Gamma_{\tau-t}^* \, .
\ee

Finally, we obtain:
\ba
P^F(\Sigma) &=& \int d\Gamma_0 \, f_0^F(\Gamma_0) \, \delta\left( \Sigma - \Sigma^F[\Gamma_t] \right) \nonumber \\
&=& \int d\Gamma_0 \, f_0^R(\Gamma_0^\dagger) \, \frac{ f_0^F(\Gamma_0) }{ f_0^R(\Gamma_0^\dagger) } \, \delta\left( \Sigma + \Sigma^R[\Gamma_t^\dagger] \right) \nonumber \\
&=& \int d\Gamma_0^\dagger \, f_0^R(\Gamma_0^\dagger) \, \frac{ f_0^F(\Gamma_\tau^{\dagger *}) }{ f_0^R(\Gamma_0^\dagger) } \, \delta\left( \Sigma + \Sigma^R[\Gamma_t^\dagger] \right) \nonumber \\
&=& \int d\Gamma_0^\dagger \, f_0^R(\Gamma_0^\dagger) \, \frac{ \rho_0^F(X_\tau^\dagger,-V_\tau^\dagger) }{ \rho_0^R(X_0^\dagger,V_0^\dagger) }
\frac{ \Pi({\bf y}_\tau^{\dagger *}, {\bf z}_\tau^{\dagger *} \vert X_\tau^\dagger ) }{ \Pi({\bf y}_0^\dagger, {\bf z}_0^\dagger \vert X_0^\dagger ) }
\, \delta\left( \Sigma + \Sigma^R[\Gamma_t^\dagger] \right) \nonumber \\
&=& \int d\Gamma_0^\dagger \, f_0^R(\Gamma_0^\dagger) \,
e^{-\Sigma^R[\Gamma_t^\dagger]}
\, \delta\left( \Sigma + \Sigma^R[\Gamma_t^\dagger] \right) \nonumber \\
\label{eq:cft-a-explicit}
&=& e^\Sigma \, P^R(-\Sigma)
\ea
using Eqs.~\ref{eq:f0FR}, \ref{eq:SigmaR-convenient} and \ref{eq:Pi-tri}-\ref{eq:conjTwin}.

Eq.~\ref{eq:cft-a-explicit} is the counterpart of Eq.~64 of the main text.







\end{document}